\documentclass[prb,twocolumn,superscriptaddress,amsmath,amssymb,floatfix]{revtex4-2}
\usepackage{graphicx}
\usepackage{bm}
\usepackage{dcolumn}
\usepackage[utf8]{inputenc}
\usepackage[T1]{fontenc}
\usepackage[hidelinks]{hyperref}

\newcommand{\kk}{\mathbf{k}}
\newcommand{\qq}{\mathbf{q}}
\newcommand{\QQ}{\mathbf{Q}}
\newcommand{\RR}{\mathbf{R}}
\newcommand{\Ub}{\bar U}

\begin{document}

\title{Charge-fluctuation hot lines and superconductivity in Ba$_{1-x}$K$_x$BiO$_3$}

\author{Xiangang Wan}
\affiliation{National Laboratory of Solid State Microstructures, School of Physics and Collaborative Innovation Center of Advanced Microstructures, Nanjing University, Nanjing 210093, China}
\author{Sergey Y. Savrasov}
\affiliation{National Laboratory of Solid State Microstructures, School of Physics and Collaborative Innovation Center of Advanced Microstructures, Nanjing University, Nanjing 210093, China}
\affiliation{Department of Physics, University of California, Davis, CA 95616, USA}

\date{\today}

\begin{abstract}
The origin of superconductivity near 30~K in Ba$_{0.6}$K$_{0.4}$BiO$_3$ has remained a long-standing puzzle. Tunneling experiments give an electron--boson coupling $\lambda\approx 0.7$--1.2, whereas density-functional calculations give $\lambda\approx 0.3$, concentrated in the breathing mode near 50--70~meV. We propose that the missing coupling of about 0.7 comes from dynamical charge fluctuations at $\QQ=(\pi,\pi,\pi)/a$, the ordering vector of the charge-disproportionated parent compound, and introduce an overdamped boson model to describe them. Using a tight-binding fit to the conduction band found from density-functional calculations at $x=0.4$, we show that such a boson scatters electrons strongly only on hot lines of the Fermi surface, small loops around the $\langle 111\rangle$ directions where $\kk$ and $\kk+\QQ$ are both at the Fermi level. We derive closed-form self-energies, quasiparticle residues and band reconstructions, including shadow bands and hot-line pseudogaps for long-range correlations, and show that a sum rule fixes the Fermi-surface-averaged coupling independently of the boson frequency and correlation length. Combining the LDA phonon with an overdamped mode of frequency $\omega_m\approx 25$~meV and coupling $\lambda\approx 0.7$, anisotropic Eliashberg calculations reproduce $T_c\approx 25$--30~K for correlation lengths of two to three lattice spacings and predict a gap enhanced by a factor of 3--4 on the hot loops. For longer correlation lengths an $s$-wave state competes closely with a sign-changing $T_{2g}$ state confined to the hot loops. In the random-phase approximation for an extended Hubbard model, the boson emerges naturally near a charge instability at $R$, but purely electronic fluctuations are too weak and too long-ranged; coupling to the breathing-mode lattice distortion is one possible way to strengthen them.
\end{abstract}

\maketitle

\section{Introduction}

The bismuthate perovskites were among the first oxide superconductors with transition temperatures well above those of conventional intermetallics. Superconductivity in BaPb$_{1-x}$Bi$_x$O$_3$ was reported in 1975~\cite{Sleight1975}, and potassium doping of BaBiO$_3$ raised $T_c$ to about 30~K in cubic Ba$_{1-x}$K$_x$BiO$_3$ (BKBO) near $x=0.4$~\cite{Cava1988,Mattheiss1988b,Hinks1988}. The parent compound BaBiO$_3$ is a charge-ordered insulator in which Bi$^{3+}$/Bi$^{5+}$ disproportionation freezes a breathing distortion of the BiO$_6$ octahedra, together with octahedral tilts~\cite{Cox1979}. This disproportionation motivated early negative-$U$ and local-pair pictures of the bismuthates~\cite{Rice1981}. Doping suppresses long-range charge order, and the cubic metal that emerges has no magnetic moments, a single conduction band of Bi-$6s$/O-$2p$ character, and a three-dimensional Fermi surface.

Tunneling spectroscopy provided the first strong evidence that pairing in BKBO is mediated by bosons in the phonon energy range. Point-contact junctions revealed reproducible structure above the gap, and McMillan--Rowell inversion yielded an Eliashberg function $\alpha^2F(\omega)$ resembling the phonon density of states~\cite{Huang1990}. The reproducibility of these features was questioned~\cite{Sharifi1991}, but later measurements in high magnetic fields isolated three stable peaks and recovered spectral weight over 15--65~meV, in correspondence with the phonon bands seen by inelastic neutron scattering~\cite{Samuely1993,Loong1992}. A sizable oxygen isotope effect, $\alpha\approx 0.4$, has been reported~\cite{Loong1991}, in contrast to the small value found in early measurements~\cite{Batlogg1988}. More recently, angle-resolved photoemission found an isotropic gap with $2\Delta/k_BT_c=3.51$ and a dispersion anomaly near 50~meV, from which an electron--phonon coupling $\lambda\approx 1.3$ was extracted~\cite{Wen2018}.

The tunneling spectrum is the most direct constraint on theory. McMillan--Rowell inversions give $\lambda\approx 1.2$~\cite{Huang1990} and $\lambda=0.7$--1.0 with $\mu^*=0$--0.07, reproducing $T_c\approx 31$~K~\cite{Samuely1993}. Inserting these values into the Allen--Dynes formula~\cite{AllenDynes1975} gives a logarithmic average frequency $\omega_{\log}\approx 30$--40~meV, below the breathing-mode energy, and the measured spectra show weight spread over 10--70~meV rather than concentrated in the highest oxygen modes. We therefore take $\lambda\approx 1$ as the target. Optical data are harder to reconcile: analyses of the infrared conductivity~\cite{Puchkov1994} suggest $\lambda\le 0.2$ in the normal state~\cite{Marsiglio1996,Kaufmann1998}, while the superconducting-state reflectivity requires $\lambda\approx 0.7$~\cite{Kaufmann1998}.

First-principles calculations have not reproduced this spectrum. Within the local density approximation (LDA), the conduction band is well described~\cite{Mattheiss1988}, but the electron--phonon coupling is weak. Early estimates ranged from $\lambda\approx 0.4$ in LDA~\cite{Hamada1989} to $\lambda>1$ in tight-binding models for $x<0.3$~\cite{Shirai1990}, and the full linear-response calculation of Meregalli and Savrasov gave a total $\lambda=0.34$ for Ba$_{0.6}$K$_{0.4}$BiO$_3$~\cite{Meregalli1998}. In that calculation the coupling is concentrated in the bond-stretching and breathing branch at 65--75~meV, which softens by only about 10\% toward the zone-boundary points $R$ and $M$, where its mode coupling peaks at $\lambda\approx 0.3$. LDA therefore misses the measured coupling by a factor of about three, leaving $\lambda\approx 0.7$ unaccounted for. With $\lambda\approx 0.3$ at these frequencies, McMillan's formula gives $T_c$ of only a few kelvin, from about 1~K for $\mu^*=0.1$ to about 4~K for $\mu^*=0.05$.

Two lines of work have sought to close this gap. The first emphasizes nonlocal correlations. Hybrid-functional calculations reproduce the breathing distortion and gap of BaBiO$_3$ and predict polaronic hole trapping upon doping~\cite{Franchini2009}. Yin, Kutepov and Kotliar argued that LDA systematically underestimates electron--phonon coupling near a metal--insulator transition, and found a large $GW$ enhancement of the breathing-mode coupling~\cite{Yin2013}. Li \textit{et al.} extended this to all wave vectors with $GW$ perturbation theory and reproduced $T_c$ and its doping dependence~\cite{Li2019}. These calculations rescale the coupling to the existing phonon branches but leave the phonon spectrum essentially as in LDA. They therefore concentrate the enhanced coupling at 50--75~meV, whereas tunneling shows coupling spread over a broad range of lower frequencies. The second line invokes lattice polarons, which narrow the band and enhance the effective density of states and coupling~\cite{Alexandrov1996,Bischofs2002,Nourafkan2012,Derimow2014}. Photoemission, however, finds a bandwidth wider than LDA rather than narrower~\cite{Wen2018}, and tunneling and photoemission both show a full, BCS-like gap with no sign of preformed pairs.

None of these approaches identifies the missing coupling of about 0.7 beyond the LDA value. Here we propose that the missing ingredient is the dynamical charge fluctuation at the ordering vector of the parent compound, $\QQ=(\pi,\pi,\pi)/a$. Near a charge instability, the screened interaction becomes strongly peaked at $\QQ$ and overdamped by particle--hole pairs, a situation studied extensively for incommensurate charge order~\cite{Castellani1995,Perali1996} and for antiferromagnetic fluctuations in the spin-fermion model~\cite{Millis1990,Abanov2003}. In three-dimensional BKBO the scattering is concentrated on lines of the Fermi surface where $\kk$ and $\kk+\QQ$ both lie at the Fermi level. On these hot lines the low-energy coupling is strong, while the Fermi-surface average remains constrained by a simple sum rule.

These ideas have a long history in the copper oxides. In the nearly antiferromagnetic Fermi-liquid picture, Millis, Monien and Pines introduced a phenomenological dynamical susceptibility peaked at $\QQ=(\pi,\pi)$ and overdamped by particle--hole excitations~\cite{Millis1990}. Monthoux, Balatsky and Pines used it as a pairing interaction and obtained $d$-wave superconductivity with $T_c$ in the observed range~\cite{Monthoux1991,Monthoux1993}, building on the earlier finding that antiferromagnetic fluctuations favor $d$-wave pairing in the Hubbard model~\cite{Scalapino1986}; the spin-fluctuation mechanism is reviewed in Ref.~\cite{Scalapino2012}. Hlubina and Rice showed that scattering by such fluctuations is concentrated at hot spots, where $\kk$ and $\kk+\QQ$ both lie on the Fermi surface, while the cold regions short-circuit transport~\cite{Hlubina1995}. Close to the ordered phase, precursor effects at the hot spots open a pseudogap once the correlation length exceeds a scale set by the Fermi velocity and the coupling~\cite{Vilk1997,Schmalian1998}. Abanov, Chubukov and Schmalian turned the spin-fermion model into a quantitative theory of the quantum-critical normal state~\cite{Abanov2003} and of pairing near the instability~\cite{Abanov2001}, and Metlitski and Sachdev established by renormalization group that Fermi-liquid behavior breaks down at the hot spots~\cite{Metlitski2010}.

The charge-channel analogue was put forward by Castellani, Di Castro and Grilli, who showed that the effective interaction near an incommensurate charge-density-wave instability is singular and overdamped, producing anomalous quasiparticle scattering~\cite{Castellani1995} and strong pairing~\cite{Perali1996}. Resonant x-ray scattering later found dynamical charge-density fluctuations in cuprates with characteristic energies of a few meV over a broad range of doping~\cite{Arpaia2019}, and these have been analysed with the same framework~\cite{Caprara2017}. Momentum-selective coupling also appears in other contexts: correlations can favor forward scattering in the electron--phonon interaction of the cuprates~\cite{Kulic2000}, and spin fluctuations connecting nested Fermi-surface pockets underlie the $s_\pm$ pairing scenario in the iron pnictides~\cite{Mazin2008}. All of these theories address quasi-two-dimensional materials near magnetic or incommensurate charge order. BKBO offers a simpler setting: a three-dimensional single band, a commensurate ordering vector with $2\QQ$ a reciprocal lattice vector, and an attractive charge-channel interaction that favors a same-sign gap on $\kk$ and $\kk+\QQ$. In three dimensions the hot spots become hot lines.

In this paper we introduce an overdamped boson model for these dynamical charge fluctuations and develop a phenomenological theory of the electrons coupled to it, using a tight-binding fit to the BKBO conduction band found from density-functional calculations. Section~\ref{sec:normal} introduces the model interaction and computes the normal-state self-energy, quasiparticle renormalization and band structure. Section~\ref{sec:sc} solves the anisotropic Eliashberg equations with this boson and a weak background phonon coupling, and compares the results with the tunneling $\alpha^2F(\omega)$ and $T_c$. Section~\ref{sec:rpa} asks whether such a boson can arise microscopically, using the random-phase approximation for an extended Hubbard model, and discusses the breathing-mode electron--phonon coupling as one possible additional ingredient. Section~\ref{sec:concl} presents our conclusion.

\section{Normal state}
\label{sec:normal}

\subsection{Model}

\begin{figure*}[t]
\includegraphics[width=0.9\textwidth]{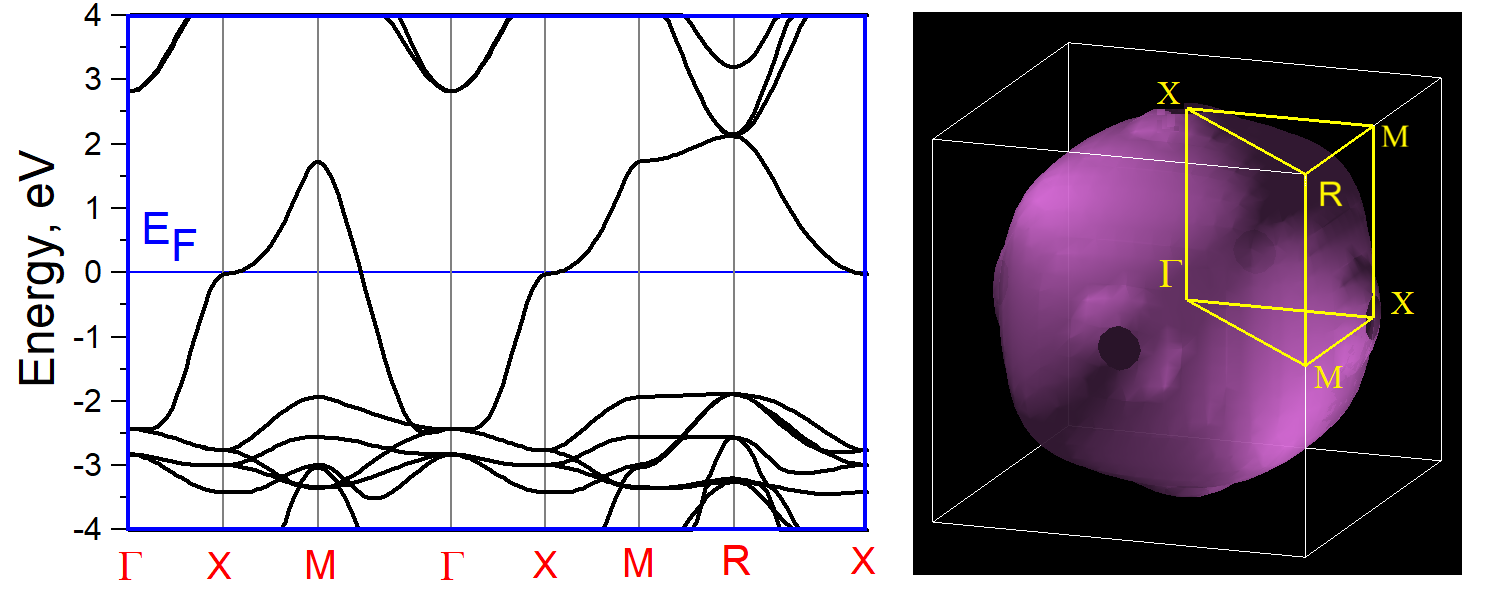}
\caption{Left: LDA band structure of cubic BaBiO$_3$ along $\Gamma$--$X$--$M$--$\Gamma$--$X$--$M$--$R$--$X$, with energies relative to the Fermi level $E_F$ of the undoped compound ($n=1$). The single band crossing $E_F$, of Bi-$6s$/O-$2p$ antibonding character, is the one fitted in Appendix~\ref{app:tb}. Right: the Fermi surface of the undoped compound inside the simple-cubic Brillouin zone, with the high-symmetry points $\Gamma=(0,0,0)$, $X=(\pi,0,0)$, $M=(\pi,\pi,0)$ and $R=(\pi,\pi,\pi)$ in units of $1/a$; the second part of the path runs through the equivalent points $X=(0,0,\pi)$ and $M=(\pi,0,\pi)$. The zone corner $R$ is the ordering vector $\QQ$ of the charge-ordered parent compound.}
\label{fig:lda}
\end{figure*}

We describe the conduction electrons of BKBO by a single band $\varepsilon_\kk$ on the simple-cubic lattice with lattice constant $a$, obtained from a tight-binding fit, with 14 shells of hopping parameters, to the band found from density-functional calculations in the LDA (Appendix~\ref{app:tb}). All results are for $n=0.6$ electrons per Bi site, corresponding to $x=0.4$; the chemical potential is $\mu=-0.69$~eV relative to half filling and the density of states per spin is $N_\sigma(0)=0.21$~eV$^{-1}$. We write $\xi_\kk=\varepsilon_\kk-\mu$. Figure~\ref{fig:lda} shows the LDA band structure and the Fermi surface of the undoped compound. At the filling $n=0.6$ the Fermi surface is a single closed pocket around $\Gamma$ that does not touch the zone boundary; it is shown in Fig.~\ref{fig:hot}.

The electrons couple to a charge fluctuation peaked at the ordering vector of the parent compound, $\QQ=(\pi,\pi,\pi)/a$, the $R$ point. We take the retarded effective interaction in the relaxational form, which defines the overdamped boson model used throughout this paper,
\begin{equation}
W(\qq,\omega)=-\frac{V(\qq)}{\omega_m-i\omega},\qquad W(\qq,i\nu_l)=-\frac{V(\qq)}{\omega_m+|\nu_l|},
\label{eq:W}
\end{equation}
where $V(\qq)\ge 0$ is peaked at $\QQ$ and $\omega_m$ is the characteristic frequency of the overdamped mode. The negative sign makes the interaction attractive in the charge channel, which is required for causality of the one-loop self-energy below. The corresponding boson spectral function is
\begin{equation}
B(\qq,\omega)=-\frac{1}{\pi}\,\mathrm{Im}\,W(\qq,\omega)=\frac{V(\qq)}{\pi}\,\frac{\omega}{\omega^2+\omega_m^2},
\label{eq:B}
\end{equation}
which is broad, peaks at $\omega=\omega_m$ and decays as $1/\omega$. Equation~\eqref{eq:W} is exactly the form obtained in the random-phase approximation near a charge instability, where the real part of the bare susceptibility at $\QQ$ is nearly frequency independent and its imaginary part is linear in $\omega$ from Landau damping (Sec.~\ref{sec:rpa}). This parallels the phenomenological spin susceptibility of Millis, Monien and Pines~\cite{Millis1990}.

We characterize the strength of the interaction by its Brillouin-zone average at zero frequency,
\begin{equation}
\Ub=\int\frac{d^3q}{(2\pi)^3}\,|W(\qq,0)|=\frac{1}{\omega_m}\int\frac{d^3q}{(2\pi)^3}\,V(\qq),
\label{eq:Ubar}
\end{equation}
where here and below we set the lattice constant $a=1$, so that the integral runs over a Brillouin zone of volume $(2\pi)^3$ and $\Ub$ is the zone-averaged static interaction per unit cell. To one-loop order the electron self-energy is
\begin{equation}
\begin{split}
\Sigma(\kk,i\omega_n)=&-T\sum_{n'}\int\frac{d^3q}{(2\pi)^3}\\
&\times W(\qq,i\omega_n-i\omega_{n'})\,G_0(\kk+\qq,i\omega_{n'}),
\end{split}
\label{eq:Sigma}
\end{equation}
with $G_0(\kk,i\omega_n)=1/(i\omega_n-\xi_\kk)$. We first take $V(\qq)$ to be a $\delta$-function at $\QQ$, which gives closed-form results, and then spread it over a finite width $\Delta q$.

\subsection{Self-energy for a sharp peak at $\QQ$}

For $V(\qq)=(2\pi)^3\Ub\omega_m\delta(\qq-\QQ)$, the momentum integral in Eq.~\eqref{eq:Sigma} collapses and an electron at $\kk$ scatters only to $\kk+\QQ$. Because $2\QQ$ is a reciprocal lattice vector, $\kk$ and $\kk+\QQ$ form a closed pair. Performing the frequency sum and continuing to real frequency at $T=0$ gives the retarded self-energy in closed form,
\begin{align}
\mathrm{Re}\,\Sigma(\kk,\omega)&=\Ub\omega_m\,\frac{\dfrac{u}{\pi}\ln\dfrac{|u|}{\omega_m}-\dfrac{\omega_m}{2}\,\mathrm{sgn}\,\xi_{\kk+\QQ}}{u^2+\omega_m^2},\label{eq:ReS}\\
\mathrm{Im}\,\Sigma(\kk,\omega)&=-\Ub\omega_m\,\frac{|u|}{u^2+\omega_m^2}\,\theta\!\left(\xi_{\kk+\QQ}\,u\right),\label{eq:ImS}
\end{align}
with $u=\omega-\xi_{\kk+\QQ}$ and $\omega$ measured from $\mu$. The self-energy at $\kk$ depends on $\kk$ only through the energy $\xi_{\kk+\QQ}$ of its partner state, which leads to four general properties.

First, damping sets in only at $\omega=\xi_{\kk+\QQ}$, on the side of the Fermi level where the partner state is empty for particles or filled for holes. $\mathrm{Im}\,\Sigma$ then rises linearly, reaches $-\Ub/2$ at $|u|=\omega_m$, and decays as $1/|u|$. Second, $\mathrm{Re}\,\Sigma$ has a logarithmic kink at the same energy, required by causality: through the Kramers--Kronig relation, the abrupt onset of $\mathrm{Im}\,\Sigma$ produces a term proportional to $u\ln|u|$ in $\mathrm{Re}\,\Sigma$. Third, for $|\xi_{\kk+\QQ}|\gg\omega_m$ Eq.~\eqref{eq:ReS} reduces to $\mathrm{Re}\,\Sigma\approx\Delta_{\rm eff}^2/(\omega-\xi_{\kk+\QQ})$ with $\Delta_{\rm eff}^2=(\Ub\omega_m/\pi)\ln(|u|/\omega_m)$, the self-energy of a band coupled to its $\QQ$-shifted copy by a weak, logarithmically modulated gap; this drives the band reconstruction of Sec.~\ref{sec:bands}. Fourth, the quasiparticle mass renormalization at the Fermi level is
\begin{equation}
\begin{split}
\frac{m^*_\kk}{m}-1&=-\frac{\partial\,\mathrm{Re}\,\Sigma}{\partial\omega}\bigg|_{\omega=0}=\frac{\Ub}{\pi\omega_m}\,F\!\left(\frac{|\xi_{\kk+\QQ}|}{\omega_m}\right),\\
F(x)&=\frac{(x^2-1)\ln x-(x^2+1)+\pi x}{(x^2+1)^2}.
\end{split}
\label{eq:mass}
\end{equation}
The function $F(x)$ is positive for all $x$, so the mass is enhanced everywhere on the Fermi surface. For $x\gg 1$, $F\approx(\ln x-1)/x^2$ and the enhancement is weak. For $x\ll 1$, $F\approx\ln(1/x)-1$, which diverges logarithmically where $\xi_{\kk+\QQ}\to 0$.

\begin{figure}[t]
\includegraphics[width=\columnwidth]{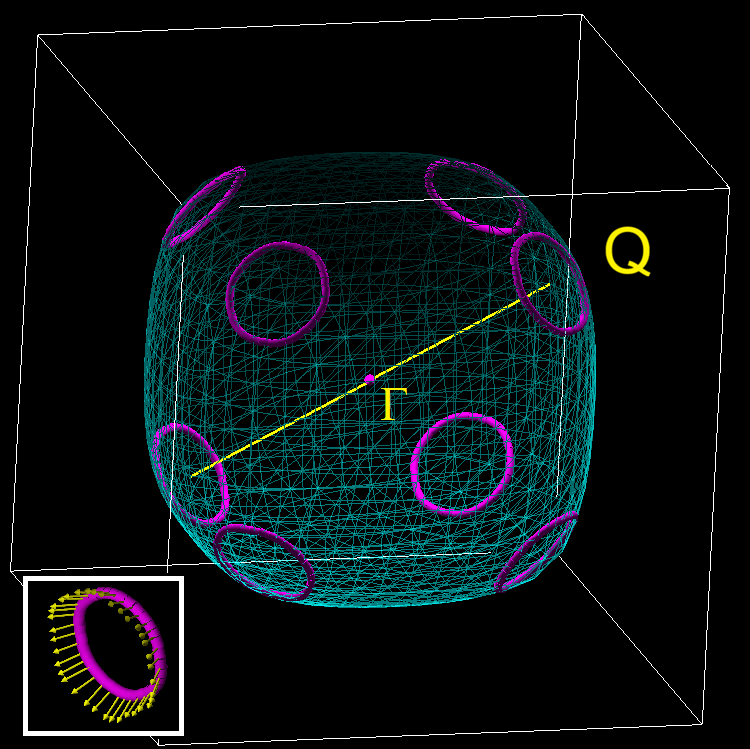}
\caption{Hot lines on the Fermi surface of Ba$_{0.6}$K$_{0.4}$BiO$_3$ ($n=0.6$). Cyan mesh: the Fermi surface from the tight-binding fit to the density-functional band, inside the simple-cubic Brillouin zone (white cube). Magenta: the hot lines, eight small loops around the $\langle 111\rangle$ directions along which the Fermi surface intersects its own copy shifted by $\QQ$, so that $\xi_\kk=\xi_{\kk+\QQ}=0$. Yellow: the vector $\QQ=(\pi,\pi,\pi)/a$, drawn through $\Gamma$, which maps each loop onto the loop around the opposite $\langle 111\rangle$ direction. Inset: one hot loop with the Fermi velocities $\mathbf{v}_\kk$ (yellow arrows) at points along it.}
\label{fig:hot}
\end{figure}

The points where both $\xi_\kk$ and $\xi_{\kk+\QQ}$ vanish form the hot lines. For our band at $n=0.6$ they are eight small closed loops on the Fermi surface, each about $12^\circ$ from a $\langle 111\rangle$ direction; one passes through $\kk\approx(0.43,0.43,0.65)\pi/a$. Inside each loop $\xi_{\kk+\QQ}<0$, and on the rest of the Fermi surface $\xi_{\kk+\QQ}>0$, reaching 2.9~eV along $\langle 100\rangle$. Figure~\ref{fig:hot} shows the hot loops on the Fermi surface, the vector $\QQ$ connecting a loop to its partner, and the Fermi velocities along a loop.

\begin{figure}[t]
\includegraphics[width=\columnwidth]{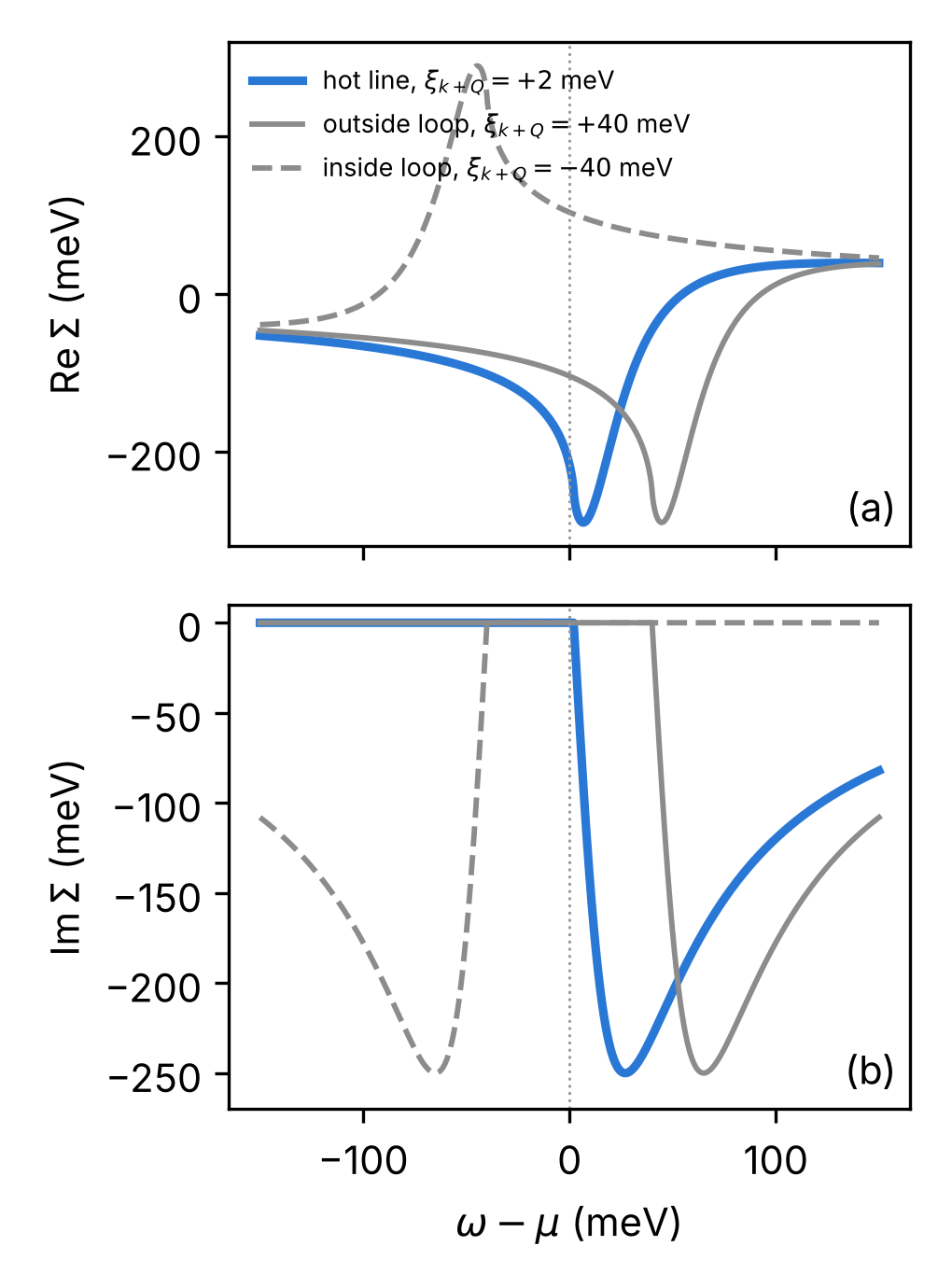}
\caption{Retarded self-energy in the $\delta$-function approximation, Eqs.~\eqref{eq:ReS}--\eqref{eq:ImS}, at three Fermi-surface points for $\Ub=0.5$~eV and $\omega_m=25$~meV: near a hot line ($\xi_{\kk+\QQ}=+2$~meV), outside a hot loop ($+40$~meV) and inside one ($-40$~meV). (a) Real part; (b) imaginary part. The dotted line marks the Fermi level. In this approximation each $\kk$ scatters into a single partner state, so the threshold onsets and the linear rise of $\mathrm{Im}\,\Sigma$ are not of Fermi-liquid form; a finite momentum width of the boson restores $\mathrm{Im}\,\Sigma\propto\omega^2$ for $|\omega|\lesssim\omega_m$ [Eq.~\eqref{eq:ImShot}].}
\label{fig:sigma}
\end{figure}

\subsection{Representative self-energy}

Figure~\ref{fig:sigma} shows the $\delta$-function results, Eqs.~\eqref{eq:ReS}--\eqref{eq:ImS}, at three Fermi-surface points, for $\Ub=0.5$~eV and $\omega_m=25$~meV, the boson frequency adopted in Sec.~\ref{sec:sc} from the tunneling spectrum. Each curve has the same shape, displaced to the partner energy $\xi_{\kk+\QQ}$. Near the hot line the damping onset and the kink in $\mathrm{Re}\,\Sigma$ sit at the Fermi level, so the quasiparticle there is strongly renormalized. Away from it, both features move to finite energy, and at the Fermi level the self-energy is weak and nearly constant. For points far from the hot lines, such as the $\langle 100\rangle$ and $\langle 110\rangle$ directions where $\xi_{\kk+\QQ}$ is 1.1--2.9~eV, the features lie entirely outside the low-energy window. Because each $\kk$ scatters into a single partner state in this approximation, the threshold onsets and the linear rise of $\mathrm{Im}\,\Sigma$ in Fig.~\ref{fig:sigma} are not of Fermi-liquid form; we show below that a finite momentum width restores $\mathrm{Im}\,\Sigma\propto\omega^2$ at $|\omega|\lesssim\omega_m$.

\subsection{Finite momentum width}

A real charge fluctuation has a finite correlation length $\xi_c$, so $V(\qq)$ is spread over a width $\Delta q=1/\xi_c$ around $\QQ$. Two forms are standard. The Ornstein--Zernike form~\cite{Ornstein1914}, $V(\qq)\propto[\xi_c^{-2}+(\qq-\QQ)^2]^{-1}$, is the one used by Millis, Monien and Pines~\cite{Millis1990} and the one produced by RPA near a continuous instability (Sec.~\ref{sec:rpa}). In three dimensions its tails decay slowly and its momentum integral must be cut off at the zone boundary. The squared Lorentzian,
\begin{equation}
V(\qq)=(2\pi)^3\,\Ub\omega_m\,\frac{\Delta q}{\pi^2\left[(\qq-\QQ)^2+\Delta q^2\right]^2},
\label{eq:Vlor}
\end{equation}
is the Fourier transform of correlations decaying as $e^{-r/\xi_c}$. It is normalized in three dimensions and reduces to the $\delta$-function above as $\Delta q\to 0$. We also consider Eq.~\eqref{eq:Vlor} truncated at $|\qq-\QQ|=5\Delta q$ and renormalized to the same $\Ub$, to test the role of the tails.

Linearizing $\xi_{\kk+\qq}\approx\xi_{\kk+\QQ}+\mathbf{v}_{\kk+\QQ}\cdot(\qq-\QQ)$ and integrating Eq.~\eqref{eq:Vlor} over the two directions perpendicular to the partner velocity gives exactly a Lorentzian distribution of partner energies, with width $\Gamma_\kk=|\mathbf{v}_{\kk+\QQ}|\Delta q$. Averaging Eq.~\eqref{eq:mass} over this distribution gives, on the hot line and without further approximation,
\begin{equation}
\frac{m^*}{m}-1\bigg|_{\text{hot line}}=\frac{\Ub}{\pi\omega_m}\,\frac{x-1-\ln x}{(x-1)^2},\qquad x=\frac{\Gamma}{\omega_m}.
\label{eq:masshot}
\end{equation}
For $\Gamma\ll\omega_m$ this reduces to $(\Ub/\pi\omega_m)[\ln(\omega_m/\Gamma)-1]$, the $\delta$-function result with $|\xi_{\kk+\QQ}|$ replaced by $\Gamma$: momentum smearing cuts off the logarithmic divergence. For $\Gamma\gg\omega_m$ it becomes $\Ub/(\pi\Gamma)$, independent of $\omega_m$. The Fermi velocity on the hot line is $|\mathbf{v}_{\kk+\QQ}|\approx 1.8$~eV$\cdot a$, so $\Gamma$ exceeds $\omega_m\approx 25$~meV unless $\xi_c\gtrsim 70a$. For any realistic correlation length the hot-line enhancement is therefore set by $\Ub/(\pi v_F\Delta q)$.

Momentum smearing also restores Fermi-liquid behavior of the damping. In the $\delta$-function limit, Eq.~\eqref{eq:ImS} gives a threshold at $|\omega|=|\xi_{\kk+\QQ}|$ away from the hot lines and, on them, $\mathrm{Im}\,\Sigma\approx-\Ub|\omega|/\omega_m$ with $\mathrm{Re}\,\Sigma\propto\omega\ln|\omega|$, the marginal-Fermi-liquid form. Both reflect the absence of phase space: an electron can decay into one final state only. Averaging Eq.~\eqref{eq:ImS} over the Lorentzian distribution of partner energies gives, on the hot line and for $|\omega|\ll\Gamma$,
\begin{equation}
\mathrm{Im}\,\Sigma_{\rm hot}(\omega)\simeq-\frac{\lambda_{\rm hot}\,\omega_m}{2}\,\ln\!\left(1+\frac{\omega^2}{\omega_m^2}\right),
\label{eq:ImShot}
\end{equation}
with $\lambda_{\rm hot}=\Ub/(\pi\Gamma)$. For $|\omega|\ll\omega_m$ this is the Fermi-liquid form, $\mathrm{Im}\,\Sigma\approx-\lambda_{\rm hot}\omega^2/(2\omega_m)$, and for $\omega_m\ll|\omega|\ll\Gamma$ the damping grows only logarithmically. The Fermi-surface average has the same structure, since the distribution of $\xi_{\kk+\QQ}$ is smooth near zero. The hot regions are therefore a Fermi liquid with a small coherence scale set by $\omega_m\approx 25$~meV, implying a $T^2$ contribution to the resistivity only below roughly $\omega_m/k_B$ and a crossover to weaker temperature dependence above it. The coherence scale shrinks as the charge instability is approached; at the critical point itself the boson is a Landau-damped mode with dynamical exponent $z=2$, meaning that its characteristic frequency scales with the correlation length as $\omega_m\propto\xi_c^{-2}$ (Sec.~\ref{sec:rpawidth}), and in three dimensions it gives the Hertz--Millis form~\cite{Hertz1976,Millis1993} $\mathrm{Im}\,\Sigma\propto|\omega|^{3/2}$ on the hot lines.

The Fermi-surface average behaves very differently. The kernel in Eq.~\eqref{eq:mass}, viewed as a function of $\xi_{\kk+\QQ}$, has a fixed integral, $\int d\xi\,(m^*/m-1)/\Ub=1$, independent of $\omega_m$. Averaging over the Fermi surface samples this peak with the distribution of partner energies, which gives the sum rule
\begin{equation}
\begin{split}
\left\langle\frac{m^*}{m}-1\right\rangle_{\rm FS}&\simeq\Ub\,P_\QQ(0),\\ P_\QQ(0)&=\frac{\sum_\kk\delta(\xi_\kk)\,\delta(\xi_{\kk+\QQ})}{\sum_\kk\delta(\xi_\kk)}.
\end{split}
\label{eq:sumrule}
\end{equation}
The nesting density $P_\QQ(0)$ is the fraction of the Fermi surface, per unit energy, whose partner also lies at the Fermi level. For our band at $n=0.6$, Monte Carlo sampling gives $P_\QQ(0)\approx 0.22$~eV$^{-1}$, consistent with $J_\QQ(0,0)/N_\sigma(0)$ in Sec.~\ref{sec:rpa}. Equation~\eqref{eq:sumrule} holds as long as $\omega_m$ and $\Gamma$ are small compared with the scale on which the distribution of $\xi_{\kk+\QQ}$ varies. Because the kernel decays only as $\ln|\xi|/\xi^2$, it samples partner energies out to several tenths of an eV, where their distribution is somewhat larger than at $\xi=0$; the full numerical average below is therefore about 25\% larger than $\Ub P_\QQ(0)$.

We verified these results (Table~\ref{tab:mass}) by integrating Eq.~\eqref{eq:Sigma} numerically over the full Fermi surface with the truncated form of Eq.~\eqref{eq:Vlor}, without linearizing the band, for $\omega_m=25$~meV. Per eV of $\Ub$, the Fermi-surface average of $m^*/m-1$ decreases only from 0.27 to 0.20 as $\Delta q$ increases from 0.001 to 0.05 ($2\pi/a$), close to Eq.~\eqref{eq:sumrule}. Over the same range the hot-line value drops from 12.8 to 0.69, and the median over the Fermi surface rises from 0.018 to 0.09. Truncating the tails changes the average by less than 20\%. The total coupling is therefore fixed by $\Ub$ and the band, while $\Delta q$ only controls how it is distributed: a narrow peak concentrates it in a strong, thin hot band, and a broad one spreads it weakly over the Fermi surface.

\begin{table*}[t]
\caption{Quasiparticle residue from the $R$-point boson for $\omega_m=25$~meV and $\Ub=1$~eV, from the full Fermi-surface integral with the truncated form of Eq.~\eqref{eq:Vlor}. $Z$ at the hot line $=1/(1+\lambda_{\rm hot})$; $Z$ averaged over the Fermi surface $=\langle m^*/m\rangle^{-1}$. $\Gamma=v_F\Delta q$ on the hot line. A pseudogap opens on the hot lines when Eq.~\eqref{eq:pgcrit} is satisfied; for $\Ub=0.3$~eV ($\Delta_{\rm eff}\approx 54$~meV) this requires $\Gamma\lesssim 54$~meV.}
\label{tab:mass}
\begin{ruledtabular}
\begin{tabular}{cccccc}
\shortstack{Peak width\\$\Delta q$ ($2\pi/a$)} & \shortstack{Correlation\\length $\xi_c/a$} & \shortstack{Smearing\\$\Gamma$ (meV)} & \shortstack{$Z$ at\\hot line} & \shortstack{$Z$ averaged\\over FS} & \shortstack{Pseudogap at\\$\Ub=0.3$~eV}\\
\hline
0 ($\delta$-function) & $\infty$ & 0 & $\to 0$ (log) & 0.78 & yes\\
0.001 & 160 & 11 & 0.07 & 0.79 & yes\\
0.003 & 53 & 34 & 0.13 & 0.81 & yes\\
0.01 & 16 & 115 & 0.28 & 0.82 & no\\
0.03 & 5.3 & 344 & 0.49 & 0.83 & no\\
0.05 & 3.2 & 573 & 0.59 & 0.84 & no\\
\end{tabular}
\end{ruledtabular}
\end{table*}

Table~\ref{tab:mass} separates the two effects of the peak width. The hot-line enhancement falls by more than an order of magnitude as $\Delta q$ grows, following Eq.~\eqref{eq:masshot}, and the pseudogap of Sec.~\ref{sec:bands} is confined to the first rows, where the correlation length exceeds about 35 lattice spacings. The Fermi-surface average changes by only about 30\% over the same range, and the averaged residue $\bar Z$ stays between 0.78 and 0.84 for $\Ub=1$~eV. This weak dependence is the sum rule~\eqref{eq:sumrule} at work: narrowing the peak makes the hot lines heavier but thinner, broadening it makes a wider region moderately heavy, and the product is fixed by $\Ub$ and $P_\QQ(0)$. The residual decrease at large $\Delta q$ comes from the smeared peak sampling the distribution of $\xi_{\kk+\QQ}$ away from zero, where $P_\QQ$ is smaller. Quantities that average over the Fermi surface, such as the specific-heat coefficient, therefore constrain $\Ub$ but not $\xi_c$, while momentum-resolved probes of the hot lines are needed to determine $\xi_c$.

\subsection{Band structure: shadow bands and pseudogaps}
\label{sec:bands}

Neglecting the damping, the quasiparticle bands are the poles of the interacting Green function,
\begin{equation}
E=\xi_\kk+\mathrm{Re}\,\Sigma(\kk,E),\qquad Z_\kk=\left[1-\partial_E\,\mathrm{Re}\,\Sigma(\kk,E)\right]^{-1}.
\label{eq:poles}
\end{equation}
Because $\mathrm{Re}\,\Sigma\approx\Delta_{\rm eff}^2/(E-\xi_{\kk+\QQ})$ away from the partner energy, the band is affected only where it approaches its own copy shifted by $\QQ$, that is, near the surface $\xi_\kk=\xi_{\kk+\QQ}$ in the Brillouin zone. Far from it the bands are shifted by about $\Delta_{\rm eff}^2/(\xi_\kk-\xi_{\kk+\QQ})$, less than 1~meV for realistic $\Ub$, and $Z\approx 1$. Near a crossing the behavior depends on the coupling. For a $\delta$-function peak, Eq.~\eqref{eq:poles} has a single solution at the crossing below a threshold $\Ub_c\approx 52\,\omega_m$, giving a kink and reduced weight. The threshold follows from requiring three solutions at the crossing, $\Ub_c/\omega_m=\pi\min_y y(y^2+1)/(y\ln y-\pi/2)\approx 51.8$; the static term in Eq.~\eqref{eq:ReS} raises it well above the value $\pi\min_y(y^2+1)/\ln y\approx 22.6$ obtained without that term. Below the threshold, at $\Ub=0.5$~eV and $\omega_m=15$~meV, $Z\approx 0.65$ at the crossing on the $\Gamma$--$R$ line. Above $\Ub_c$ the band splits into two branches with an anticrossing, about 0.18~eV wide for $\Ub=1$~eV, and a third solution with negative residue that is strongly damped once $\mathrm{Im}\,\Sigma$ is restored. A weak shadow of the $\QQ$-shifted band appears near each crossing. In the limit $\omega_m\to 0$ the mode becomes static and the result approaches a fully folded band with real gaps, as in a commensurate charge-density-wave state.

\begin{figure*}[t]
\includegraphics[width=\textwidth]{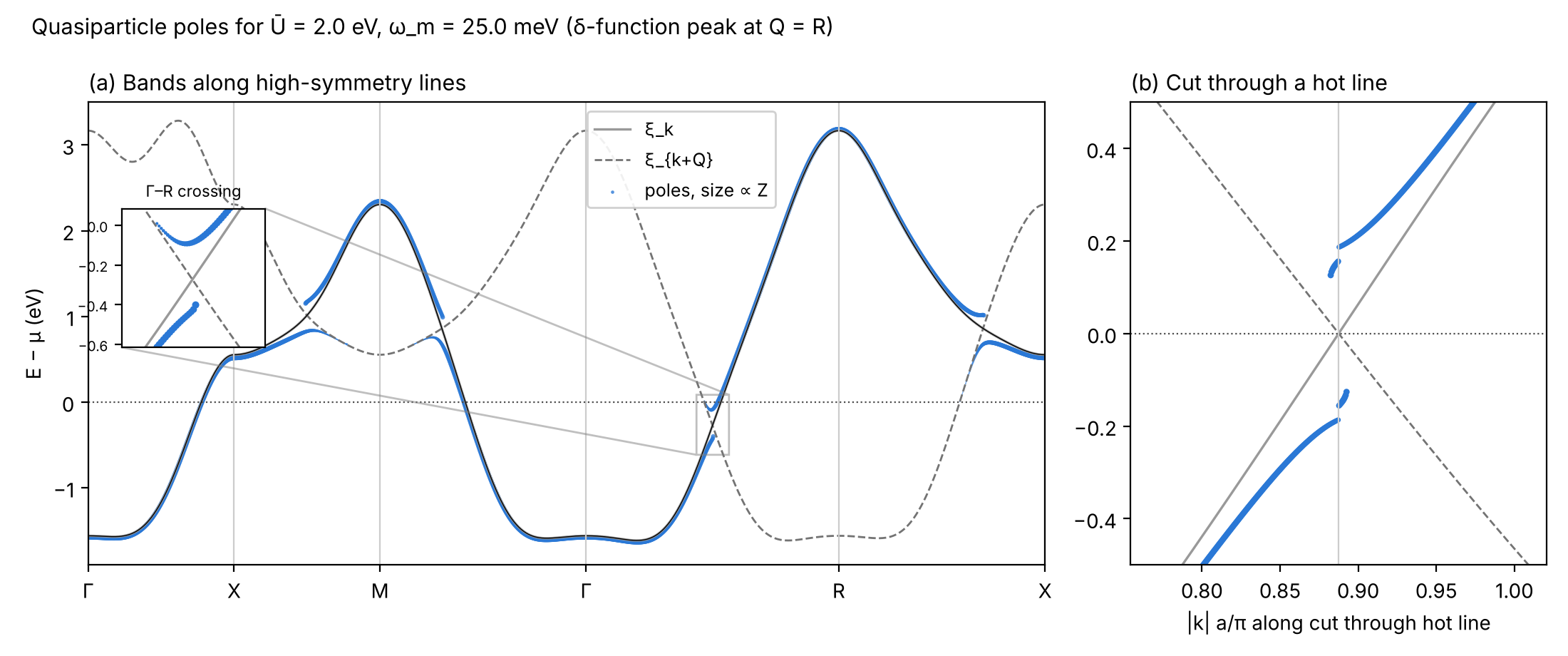}
\caption{Quasiparticle poles of Eq.~\eqref{eq:poles} for a $\delta$-function peak at $\QQ=R$, $\Ub=2$~eV, $\omega_m=25$~meV. Gray solid: bare band $\xi_\kk$; gray dashed: $\QQ$-shifted band $\xi_{\kk+\QQ}$; blue dots: poles with size proportional to $Z_\kk$ (poles with negative residue omitted). (a) High-symmetry lines, with an inset at the $\Gamma$--$R$ crossing. (b) Radial cut at $\phi=45^\circ$, $\theta=42.7^\circ$ through a hot line; the vertical line marks the bare $k_F$.}
\label{fig:bands}
\end{figure*}

Figure~\ref{fig:bands} illustrates this for $\Ub=2$~eV and $\omega_m=25$~meV, above the threshold $\Ub_c\approx 52\,\omega_m\approx 1.3$~eV. Along the high-symmetry lines [panel (a)], the poles follow the bare band $\xi_\kk$ except near its crossings with the $\QQ$-shifted band $\xi_{\kk+\QQ}$, which occur on the $X$--$M$, $M$--$\Gamma$, $\Gamma$--$R$ and $R$--$X$ lines. At each crossing the band splits and a weak shadow branch follows $\xi_{\kk+\QQ}$ over a short range, as in the inset for the $\Gamma$--$R$ crossing 0.25~eV below $\mu$. Panel (b) shows a radial cut through a hot line, where the crossing lies at the Fermi level and a gap of about 0.35~eV opens. The poles are also shifted by roughly $+0.1$ to $+0.2$~eV away from the crossings; this is the static shift that is absorbed into $\mu$ when the electron count is fixed (Sec.~\ref{sec:fs}).

On the hot lines the crossing occurs at the Fermi level. For a $\delta$-function peak the static self-energy then jumps across the hot line,
\begin{equation}
\mathrm{Re}\,\Sigma(\kk,0)\to-\frac{\Ub}{2}\,\mathrm{sgn}\,\xi_{\kk+\QQ},\qquad|\xi_{\kk+\QQ}|\ll\omega_m,
\label{eq:step}
\end{equation}
reflecting the jump in the occupation of the single partner state $\kk+\QQ$, from filled to empty, as $\kk$ crosses the hot line. The quasiparticle pole therefore jumps across the Fermi level at $k_F$ instead of passing through it, and a pseudogap opens along the hot line. For $\omega_m=25$~meV the pole sits at $-25$ and $+25$~meV on either side of $k_F$ for $\Ub=0.1$~eV, and at $\pm 54$~meV for $\Ub=0.3$~eV; the gap scales as $\Ub/2$ for $\Ub\ll\omega_m$ and as $\Delta_{\rm eff}$ for $\Ub\gg\omega_m$.

This pseudogap is a property of long-range correlations. With a finite width $\Delta q$ each $\kk$ couples to a continuum of partners whose energies spread over $\Gamma=v_F\Delta q$, the occupation step in Eq.~\eqref{eq:step} is averaged out, and $\mathrm{Re}\,\Sigma(\kk,0)$ varies smoothly on the scale $\Delta q$. A smooth momentum dependence renormalizes the Fermi velocity but opens no gap. The pseudogap survives only if
\begin{equation}
\Delta_{\rm eff}\gtrsim v_F\,\Delta q\quad\Longleftrightarrow\quad\xi_c\gtrsim\frac{v_F}{\Delta_{\rm eff}},
\label{eq:pgcrit}
\end{equation}
the three-dimensional counterpart of the criterion for precursor pseudogaps near antiferromagnetic hot spots~\cite{Vilk1997,Schmalian1998}. For $\Ub=0.3$~eV and $\omega_m=25$~meV this requires $\xi_c\gtrsim 35a$. For the correlation lengths of a few lattice spacings that we infer in Sec.~\ref{sec:sc}, the hot lines host heavy quasiparticles rather than a pseudogap.

\subsection{Fermi surface, hot lines and mass enhancement}
\label{sec:fs}

\begin{figure*}[t]
\includegraphics[width=0.85\textwidth]{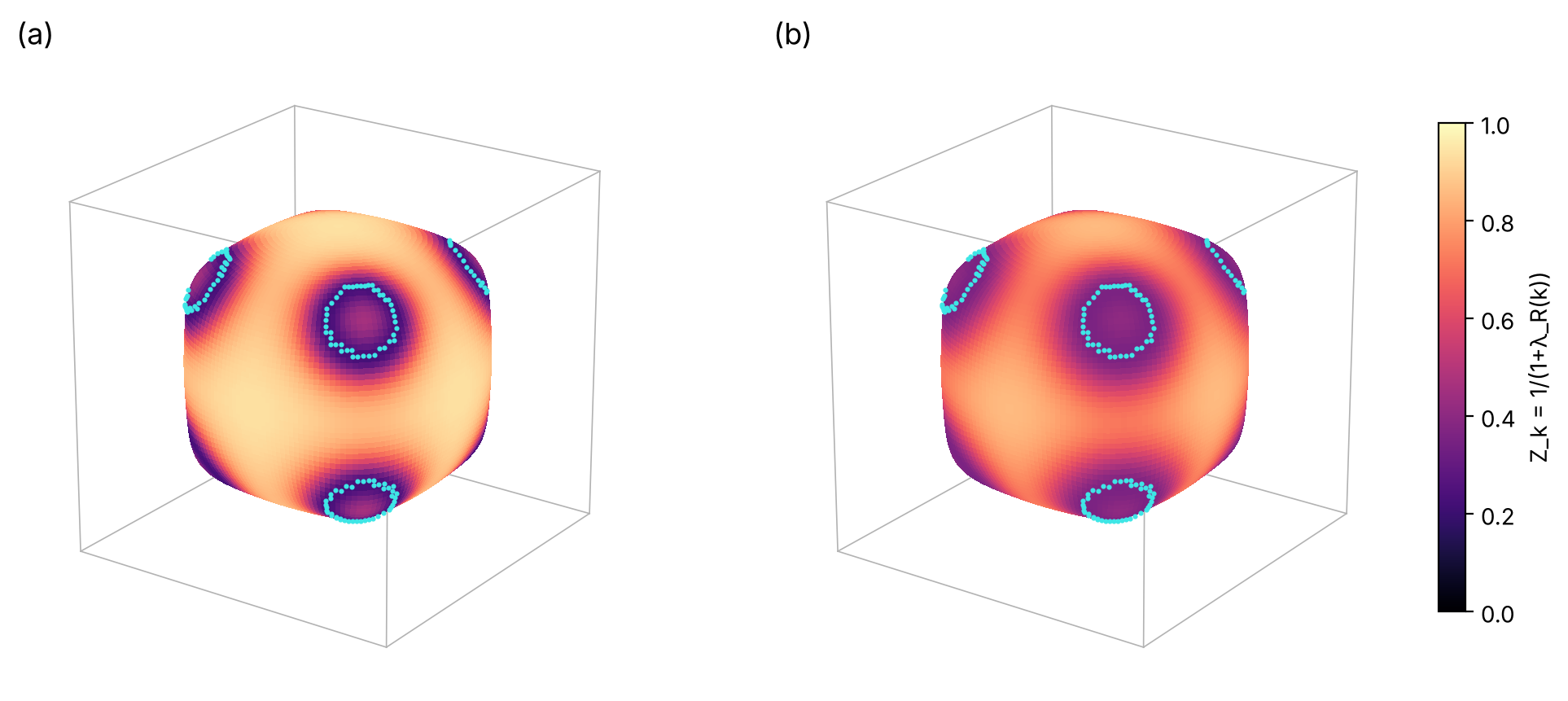}
\caption{Fermi surface at $n=0.6$ colored by $Z_\kk=1/[1+\lambda_R(\kk)]$ for the squared-Lorentzian interaction~\eqref{eq:Vlor}, with the coupling normalized to $\langle\lambda_R\rangle=0.70$. (a) $\Delta q=0.03$ ($2\pi/a$), $\xi_c\approx 5a$; (b) $\Delta q=0.07$ ($2\pi/a$), $\xi_c\approx 2.3a$. $Z_\kk$ ranges from 0.21 to 0.94 in (a) and from 0.35 to 0.86 in (b). Cyan: hot lines $\xi_\kk=\xi_{\kk+\QQ}=0$ on the visible side. The cube is the Brillouin-zone boundary at $\pm\pi/a$; the view is close to the $[111]$ direction.}
\label{fig:fsZ}
\end{figure*}

The quasiparticle residue on the Fermi surface, $Z_\kk=(m^*_\kk/m)^{-1}$ from Eq.~\eqref{eq:mass} or Eq.~\eqref{eq:masshot}, maps the hot lines directly. Figure~\ref{fig:fsZ} shows the $n=0.6$ Fermi surface colored by $Z_\kk=1/[1+\lambda_R(\kk)]$ for a narrow ($\xi_c\approx 5a$) and a broader ($\xi_c\approx 2.3a$) peak, with the hot lines marked. For a narrow peak the renormalization is confined to thin bands around the eight hot loops near $\langle 111\rangle$, where the pocket bulges toward the $R$ corners of the zone. Inside each loop, where $\xi_{\kk+\QQ}\approx -0.5$~eV at the $\langle 111\rangle$ direction itself, the enhancement is already much weaker, and on the flat faces of the pocket toward $X$, along $\langle 100\rangle$ and $\langle 110\rangle$, $Z_\kk\approx 1$. Because $F(x)>0$ in Eq.~\eqref{eq:mass}, the mass is enhanced everywhere, but the enhancement falls off as $\omega_m^2\ln(|\xi_{\kk+\QQ}|/\omega_m)/\xi_{\kk+\QQ}^2$ away from the hot lines.

As $\Delta q$ grows, the hot bands widen and weaken. Normalizing the coupling so that its Fermi-surface average is 0.70 (the value adopted in Sec.~\ref{sec:sc}), the local coupling $\lambda_R(\kk)=m^*_\kk/m-1$ ranges from 0.12 on the cold faces to 2.4 on the hot loops for $\Delta q=0.05$ ($2\pi/a$), and from 0.24 to 1.4 for $\Delta q=0.1$ ($2\pi/a$). The maximum always lies within about $10^\circ$ of $\langle 111\rangle$ and the minimum along $\langle 100\rangle$. The sum rule~\eqref{eq:sumrule} fixes the average while the anisotropy is controlled by $\xi_c$, so that a narrow peak gives a few very heavy quasiparticles and a broad peak gives moderately heavy quasiparticles over a large part of the Fermi surface.

The momentum dependence of $\Sigma$ also renormalizes the Fermi velocity. Since $\Sigma$ depends on $\kk$ only through $\xi_{\kk+\QQ}$, $\nabla_\kk\Sigma=-(\partial\Sigma/\partial\omega)\,\mathbf{v}_{\kk+\QQ}$, and the renormalized velocity is $\mathbf{v}^*_\kk=(\mathbf{v}_\kk-\Sigma'\mathbf{v}_{\kk+\QQ})/(1-\Sigma')$ with $\Sigma'=\partial_\omega\mathrm{Re}\,\Sigma$ at $\omega=0$. This term depends on the relative orientation of $\mathbf{v}_\kk$ and $\mathbf{v}_{\kk+\QQ}$ and is significant only near the hot lines.

Finally, the static part of $\Sigma$ shifts the band almost uniformly by about $-18$~meV per eV of $\Ub$ for $\omega_m=25$~meV, because $\xi_{\kk+\QQ}>0$ on most of the Fermi surface. Enforcing the Luttinger count at $n=0.6$ shifts $\mu$ by the same amount, which leaves the Fermi surface, the hot lines and the averages above unchanged to within the accuracy of the calculation.

\section{Superconductivity}
\label{sec:sc}

\subsection{Coupling spectrum}

We model the electron--boson coupling as the sum of two channels. The first is the conventional electron--phonon coupling computed in LDA, dominated by the bond-stretching and breathing branch~\cite{Meregalli1998}. We represent it by an isotropic Gaussian peak centered at $\Omega_b=50$~meV with standard deviation $\sigma=4$~meV (full width at half maximum 9.4~meV), comparable to the width of the measured peak in this frequency range, carrying $\lambda_b=0.3$. The second is the overdamped $R$-point boson of Sec.~\ref{sec:normal}. Its Eliashberg function follows from Eq.~\eqref{eq:B},
\begin{equation}
\alpha^2F_R(\omega)=\frac{\lambda_R}{\pi}\,\frac{\omega\,\omega_m}{\omega^2+\omega_m^2}\,\theta(\omega_c-\omega)\bigg/\left[\frac{2}{\pi}\arctan\frac{\omega_c}{\omega_m}\right],
\label{eq:a2FR}
\end{equation}
normalized so that $2\int\alpha^2F_R/\omega\,d\omega=\lambda_R$. We cut the spectrum at $\omega_c=61$~meV, the top of the measured spectrum, since the slow $1/\omega$ tail of the relaxational form would otherwise place unphysical weight at high frequency. Determining the actual falloff of the charge-fluctuation spectrum at high frequencies would require a full microscopic calculation, which is beyond the scope of the present phenomenological model; the sharp cutoff is the simplest choice consistent with the measured spectrum. The results depend only weakly on this choice: at fixed $\lambda_R$, varying $\omega_c$ between 45 and 90~meV changes $T_c$ by about $\pm 15\%$ (from 23 to 30~K for $\Delta q=0.07$ ($2\pi/a$), and from 18 to 23~K for uniform coupling), which shifts the consistent range of $\Delta q$ only slightly and leaves our conclusions unchanged. The sensitivity is weak because the coupling is set by the inverse-frequency moment $\lambda=2\int\alpha^2F(\omega)\,d\omega/\omega$, which we hold fixed: changing $\omega_c$ only redistributes this coupling in frequency. As the Allen--Dynes analysis of the Eliashberg equations shows~\cite{AllenDynes1975}, such a redistribution at fixed $\lambda$ enters $T_c$ mainly through the logarithmic average $\omega_{\log}$ in the prefactor rather than through the exponent, which is consistent with the weak dependence we find in the full Eliashberg solutions.

\begin{figure}[t]
\includegraphics[width=\columnwidth]{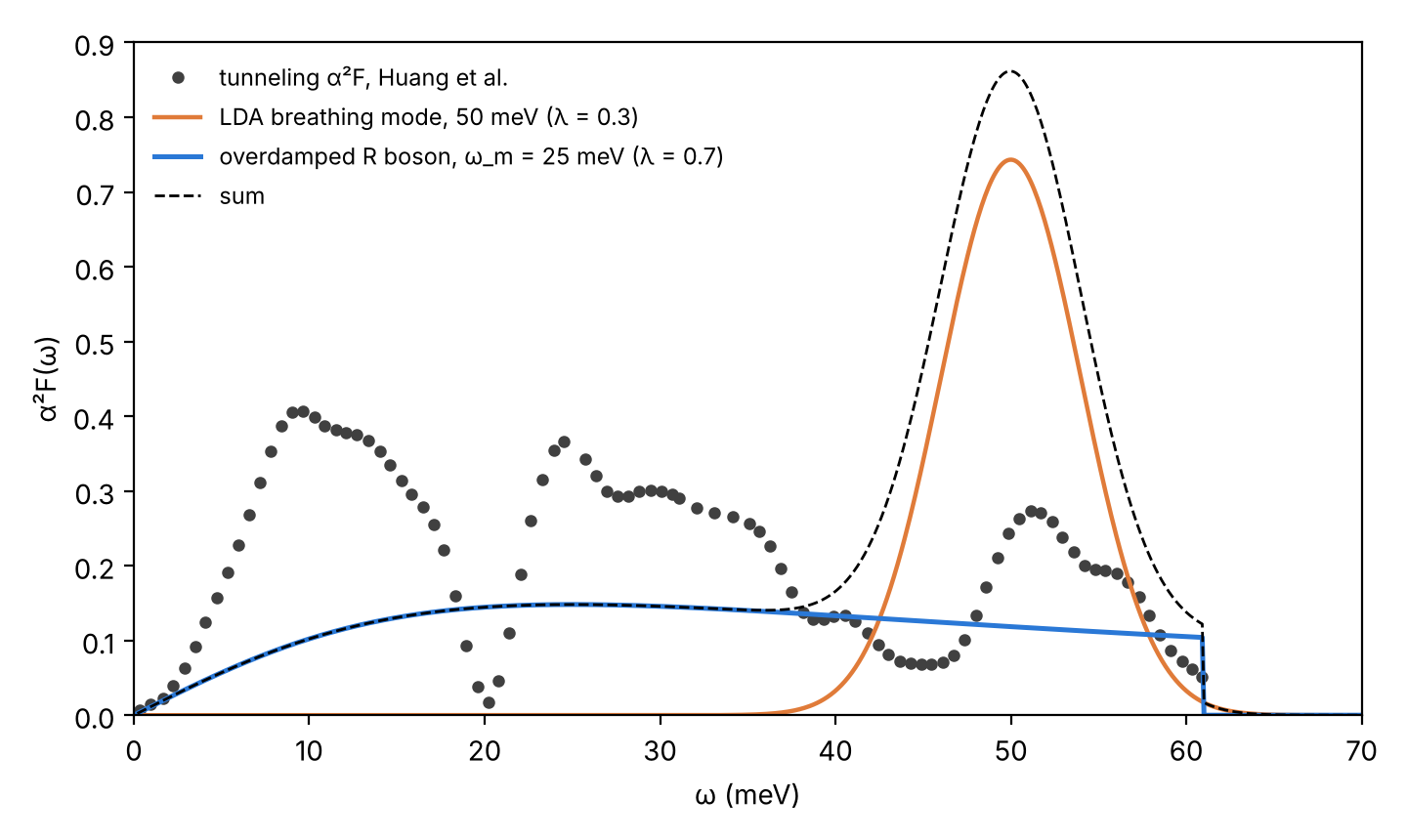}
\caption{Tunneling $\alpha^2F(\omega)$ of Ref.~\cite{Huang1990} (dots) decomposed into the LDA breathing-mode coupling, represented by a Gaussian at 50~meV with standard deviation $\sigma=4$~meV (FWHM 9.4~meV) and $\lambda_b=0.3$ (orange), and the overdamped $R$ boson of Eq.~\eqref{eq:a2FR} with $\omega_m=25$~meV and $\lambda_R=0.7$ (blue). Dashed: their sum.}
\label{fig:a2F}
\end{figure}

Tunneling gives a total coupling $\lambda\approx 0.7$--1.2~\cite{Huang1990,Samuely1993}, and we take $\lambda\approx 1$. Subtracting the LDA phonon leaves $\lambda_R\approx 0.7$ for the $R$ boson. For its frequency we take $\omega_m=25$~meV, which places the boson weight in the 10--50~meV range where the measured spectra have most of their weight and is compatible with $\omega_{\log}\approx 30$--40~meV from the Allen--Dynes analysis of the reported $\lambda$, $\mu^*$ and $T_c$; the model spectrum itself has $\omega_{\log}\approx 20$~meV, because the relaxational form keeps weight down to low frequencies. Figure~\ref{fig:a2F} compares this decomposition with the spectrum of Ref.~\cite{Huang1990}. The comparison is qualitative: the inverted spectra differ between groups and junctions, and their low-frequency part is not determined by the data (see below). It also exposes a discrepancy at high energy. The measured spectrum above 45~meV carries $\lambda\approx 0.10$, whereas the decomposition places $\lambda\approx 0.32$ there, almost all of it from the LDA peak, whose height is then nearly three times the measured $\alpha^2F$ at 50~meV. Either the LDA overestimates both the frequency and the strength of this coupling, or part of the high-energy weight is lost in the tunneling inversion. We keep $\lambda_b=0.3$ as an upper estimate of the conventional contribution. The two channels should be read as a phenomenological parameterization rather than as independent microscopic modes: the $R$ channel represents the low-energy spectral weight of the overdamped, hybridized charge--lattice mode discussed in Sec.~\ref{sec:rpa}, and the 50~meV channel the residual high-frequency phonon weight. Because the two are normalized so that their sum reproduces the measured total $\lambda\approx 1$, this hybridization does not lead to double counting.

The low-frequency part of the measured spectrum deserves a comment, because it carries a large share of $\lambda$. In the McMillan--Rowell procedure~\cite{McMillan1965}, $\alpha^2F(\omega)$ is obtained by iterating the Eliashberg equations until the computed tunneling density of states reproduces the measured conductance above the gap edge. The data are least accurate just above the gap, where the conductance varies steeply and depends on the precise value of $\Delta_0$, so the inversion is essentially insensitive to $\alpha^2F$ at frequencies below a few times $\Delta_0$. In practice the extracted spectrum is constrained there, usually by imposing a quadratic, Debye-like form $\alpha^2F\propto\omega^2$ below a cutoff that is not determined by the data; for BKBO, Samuely \textit{et al.} assumed this form below 8~meV~\cite{Samuely1993}; the choice of this region alone changes $\lambda$ by 10--20\% even in elemental superconductors~\cite{Rudin1998}. For BKBO, with $\Delta_0\approx 4$--5~meV, the spectrum below about 5~meV is therefore an extrapolation. The published curve rises roughly linearly below 3~meV and more steeply up to 7~meV. Our overdamped boson instead predicts $\alpha^2F_R\propto\omega/\omega_m$ at $\omega\ll\omega_m$, a linear onset from Landau damping rather than the $\omega^2$ of acoustic phonons. Because $\lambda$ weights the spectrum by $1/\omega$, the difference matters: below 5~meV a linear onset gives $\lambda\approx 0.35$, a quadratic one 0.17. The tunneling data cannot distinguish the two, and a linear onset is fully consistent with them.

\subsection{Eliashberg equations}

We solve the linearized Eliashberg equations on the Matsubara axis at $T=T_c$. With a coupling that depends on both momenta, they read
\begin{align}
Z(\kk,i\omega_n)&=1+\frac{\pi T}{\omega_n}\sum_{n'}\sum_{\kk'}w_{\kk'}\,\lambda(\kk,\kk',i\omega_n-i\omega_{n'})\nonumber\\
&\quad\times\mathrm{sgn}\,\omega_{n'},\nonumber\\
\Phi(\kk,i\omega_n)&=\pi T\sum_{n'}\sum_{\kk'}w_{\kk'}\Big[\lambda(\kk,\kk',i\omega_n-i\omega_{n'})\nonumber\\
&\quad-\mu^*\theta(\omega_c-|\omega_{n'}|)\Big]\frac{\Phi(\kk',i\omega_{n'})}{|\omega_{n'}|\,Z(\kk',i\omega_{n'})},
\label{eq:eliash}
\end{align}
where $w_{\kk'}$ are density-of-states weights on the Fermi surface ($\sum w_{\kk'}=1$), and $T_c$ is the highest temperature at which the gap equation has a nontrivial solution. The gap is $\Delta(\kk)=\Phi(\kk,i\pi T_c)/Z(\kk,i\pi T_c)$. The coupling is
\begin{equation}
\lambda(\kk,\kk',i\nu)=\lambda_R(\kk,\kk')\,\frac{\lambda_R(i\nu)}{\lambda_R}+\lambda_b\,\frac{\Omega_b^2}{\Omega_b^2+\nu^2},
\label{eq:lamkk}
\end{equation}
where $\lambda_R(i\nu)=2\int d\omega\,\alpha^2F_R(\omega)\,\omega/(\omega^2+\nu^2)$ carries the frequency dependence of Eq.~\eqref{eq:a2FR}, and $\lambda_R(\kk,\kk')\propto V(\kk'-\kk)$ carries the momentum structure of Eq.~\eqref{eq:Vlor}, normalized so that its Fermi-surface average equals $\lambda_R=0.70$. Because the interaction is attractive in the charge channel, the gap has the same sign at $\kk$ and $\kk+\QQ$. We discretize the Fermi surface with 800 points and use $\mu^*=0.1$ with a Matsubara cutoff of 0.25~eV unless stated otherwise. Vertex corrections are controlled by Migdal's theorem: the boson frequencies, $\omega_m\approx 25$~meV and $\Omega_b\approx 50$~meV, are small compared with the Fermi energy measured from the band bottom, about 1.6~eV, so $\omega/E_F\approx 0.02$--0.03. Strongly momentum-peaked interactions can enhance vertex corrections near hot spots, but this becomes important only for long correlation lengths, in the regime $\xi_c\gtrsim 35a$ of the pseudogap criterion~\eqref{eq:pgcrit}, far from the $\xi_c\approx 2$--3a relevant here.

A simpler hot-point form follows when the gap is assumed nonzero only on the hot lines and the partner energies are smeared by $\Gamma=v_F\Delta q$ as above. The momentum sums can then be done analytically, with the result
\begin{equation}
\Phi_n=\Ub\omega_m T\sum_{n'}\frac{\Phi_{n'}\,[\omega_m+|\omega_n-\omega_{n'}|]^{-1}}{|\omega_{n'}|Z_{n'}\,(|\omega_{n'}|Z_{n'}+\Gamma)},
\label{eq:hotpoint}
\end{equation}
and an analogous equation for $Z_n$. For $\Gamma\gg|\omega_nZ_n|$ this is the standard isotropic form with an effective coupling $\lambda_{\rm hot}=\Ub/(\pi\Gamma)$ at zero frequency, the hot-line mass enhancement of Eq.~\eqref{eq:masshot}. For $\Gamma\to 0$ it describes an isolated pair of states, $\kk$ and $\kk+\QQ$.

\subsection{Pairing by a sharp peak: superconductivity on the hot loops}
\label{sec:deltapair}

As for the normal state, the $\delta$-function limit gives the clearest picture. With $V(\qq)=(2\pi)^3\Ub\omega_m\delta(\qq-\QQ)$, the pairing interaction of Eq.~\eqref{eq:W} between states $\kk$ and $\kk'$ becomes
\begin{equation}
W(\kk'-\kk,i\nu_l)=-\frac{(2\pi)^3\,\Ub\omega_m\,\delta(\kk'-\kk-\QQ)}{\omega_m+|\nu_l|}:
\label{eq:Wdelta}
\end{equation}
it is local in momentum, connecting $\kk$ only to $\kk+\QQ$, but keeps the full frequency dependence of the overdamped boson. The momentum sum in Eq.~\eqref{eq:eliash} then collapses, and the linearized gap equation decouples into independent pairs of states, $\kk$ and $\kk+\QQ$:
\begin{equation}
\Phi_n(\kk)=\Ub\omega_m T\sum_{n'}\frac{\Phi_{n'}(\kk+\QQ)\,[\omega_m+|\omega_n-\omega_{n'}|]^{-1}}{\omega_{n'}^2Z_{n'}^2(\kk+\QQ)+\xi_{\kk+\QQ}^2},
\label{eq:deltapair}
\end{equation}
together with the same equation with $\kk$ and $\kk+\QQ$ interchanged, and analogous equations for $Z$. There is no integral over partner energies, and therefore no logarithm of the Fermi sea: each pair has its own transition temperature $T_c(\kk)$, set by the two energies $\xi_\kk$ and $\xi_{\kk+\QQ}$. It is highest when both vanish, that is, on the hot lines, where Eq.~\eqref{eq:deltapair} reduces to Eq.~\eqref{eq:hotpoint} with $\Gamma=0$.

The solution therefore develops first on the hot loops. Just below the highest $T_c$, the gap is nonzero only on the hot lines, and it spreads away from them only as the temperature is lowered further. The local $T_c$ falls off slowly: for $\Ub=0.3$~eV and $\omega_m=25$~meV it stays above 90\% of its hot-line value for $|\xi_{\kk+\QQ}|\lesssim 40$~meV on the Fermi surface and halves at $|\xi_{\kk+\QQ}|\approx 130$~meV. The gap is thus confined to a tube around the hot lines whose width in energy is several times $\pi T_c$.

\begin{figure}[t]
\includegraphics[width=\columnwidth]{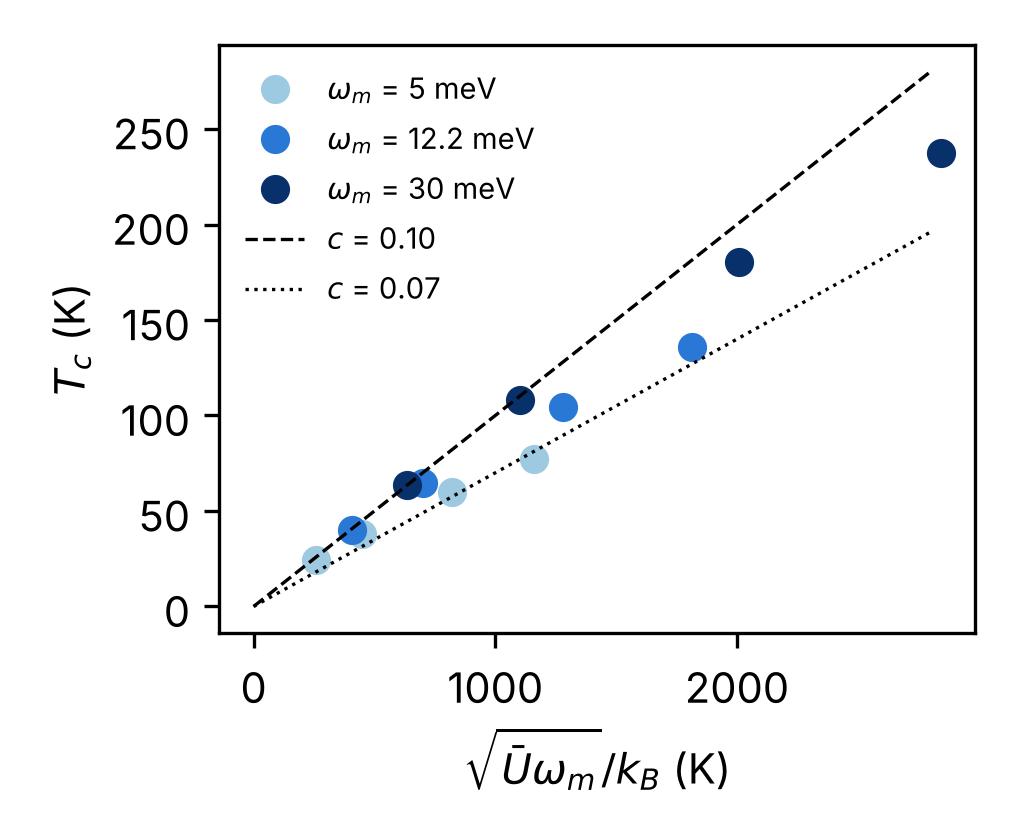}
\caption{Pairing by a $\delta$-function peak at $\QQ$. Hot-line $T_c$ from Eq.~\eqref{eq:hotpoint} with $\Gamma=0$ for $\omega_m=5$, 12.2 and 30~meV and $\Ub=0.1$--2~eV, plotted against $\sqrt{\Ub\omega_m}$; dashed and dotted lines are Eq.~\eqref{eq:Tcdelta} with $c=0.10$ and 0.07.}
\label{fig:delta}
\end{figure}

The hot-line $T_c$ has a simple, non-exponential form. In Eq.~\eqref{eq:deltapair} with $\xi_\kk=\xi_{\kk+\QQ}=0$, the terms with $n'=n$ describe exchange of a static, thermally excited boson, which acts on the electrons like elastic scattering by a frozen potential. For a gap with the same sign on $\kk$ and $\kk+\QQ$ such scattering is not pair breaking: its contributions to $\Phi$ and to $Z$ cancel exactly. This is the same cancellation that underlies Anderson's theorem for nonmagnetic impurities~\cite{Anderson1959} and the insensitivity of $T_c$ to thermally excited phonons in Eliashberg theory~\cite{Millis1988}. The remaining terms, with $|\omega_n-\omega_{n'}|\ge 2\pi T$, give a kernel of order $\Ub\omega_m/(2\pi^3T^2)$, so the condition for a nontrivial solution yields
\begin{equation}
k_BT_c\simeq c\,\sqrt{\Ub\omega_m},\qquad c\approx 0.07\text{--}0.10,
\label{eq:Tcdelta}
\end{equation}
with $c\le(2\pi^3)^{-1/2}\approx 0.13$ reduced by the mass renormalization. The numerical solutions in Fig.~\ref{fig:delta}, for $\omega_m$ from 5 to 30~meV and $\Ub$ from 0.1 to 2~eV, follow Eq.~\eqref{eq:Tcdelta} with $c$ decreasing slowly from 0.10 to 0.07 as $\Ub/\omega_m$ grows. Equation~\eqref{eq:Tcdelta} has three consequences. First, $T_c$ is set by the same energy scale, $\sqrt{\Ub\omega_m}$, as the pseudogap $\Delta_{\rm eff}$ of Sec.~\ref{sec:bands}, so pairing and the precursor band folding compete on equal footing in this limit. Second, the dependence on the coupling is a square root rather than an exponential, as in the strong-coupling limit of Eliashberg theory: a modest $\Ub\approx 0.03$~eV already gives 30~K at $\omega_m=25$~meV, while the $\Ub\approx 3$~eV implied by the tunneling coupling would give more than 200~K. Third, independent pairs at different $\kk$ have no mechanism to lock their phases. The $\delta$-function model therefore describes local pairing on the hot loops rather than a true bulk transition; a finite width $\Delta q$ couples neighboring $\kk$, restores a single $T_c$, and for $\Gamma\gg\omega_m$ returns to the conventional McMillan form with $\lambda_{\rm hot}=\Ub/(\pi\Gamma)$. The measured $T_c$ together with the tunneling coupling strength place BKBO firmly in this broad-peak regime, as we show next.

\subsection{The $R$ boson alone}

We first switch off the phonon channel. Table~\ref{tab:hotTc} gives $T_c$ from the hot-point equation~\eqref{eq:hotpoint} as a function of the coupling strength $\Ub$ and the peak width $\Delta q$, for $\omega_m=25$~meV and $\mu^*=0$.

\begin{table}[t]
\caption{$T_c$ (K) from Eq.~\eqref{eq:hotpoint}, $R$ boson alone, $\omega_m=25$~meV, $\mu^*=0$. $\Gamma=v_F\Delta q$ with $v_F=1.8$~eV$\cdot a$.}
\label{tab:hotTc}
\begin{ruledtabular}
\begin{tabular}{cccccc}
& & \multicolumn{4}{c}{$T_c$ (K)}\\ \cline{3-6}
$\Delta q$ ($2\pi/a$) & $\Gamma$ (meV) & $\Ub=0.3$~eV & $0.5$~eV & $1$~eV & $2$~eV\\
\hline
0 & 0 & 98 & 122 & 163 & 213\\
0.003 & 34 & 72 & 100 & 144 & 198\\
0.01 & 115 & 28 & 56 & 105 & 166\\
0.03 & 344 & 1.8 & 9.3 & 37 & 92\\
0.05 & 573 & 0.2 & 1.8 & 14 & 51\\
0.1 & 1150 & 0 & $<0.1$ & 1.8 & 14\\
\end{tabular}
\end{ruledtabular}
\end{table}

For realistic widths, $\Delta q\gtrsim 0.01$ ($2\pi/a$), $T_c$ depends essentially on the single combination $\lambda_{\rm hot}=\Ub/(\pi v_F\Delta q)$ and follows the McMillan form with prefactor $\omega_m/1.2$. For narrower peaks $T_c$ becomes large and insensitive to $\Ub$: each $\kk$ pairs with essentially one partner state, the regime in which Sec.~\ref{sec:bands} also predicts a pseudogap. There the pseudogap competes with superconductivity, and the values in the upper rows should not be taken literally.

The full anisotropic solution of Eq.~\eqref{eq:eliash} removes the hot-point assumption. We fix the Fermi-surface average to $\lambda_R=0.70$, the coupling missing from the LDA; by the sum rule~\eqref{eq:sumrule} this corresponds to $\Ub\approx 3$~eV, nearly independent of $\Delta q$ (from 2.5~eV for a sharp peak to 3.5~eV for $\Delta q=0.05$). The results are collected in Table~\ref{tab:aniso} together with the two-channel case. With the $R$ boson alone and the coupling spread uniformly over the Fermi surface, $T_c=6$~K. Concentrating the same average coupling around the hot lines raises $T_c$ to 7, 12, 28 and 45~K for $\Delta q=0.2$, 0.1, 0.05 and 0.03 ($2\pi/a$). This is the generic enhancement of $T_c$ by gap anisotropy at fixed average coupling.

\begin{figure*}[t]
\includegraphics[width=\textwidth]{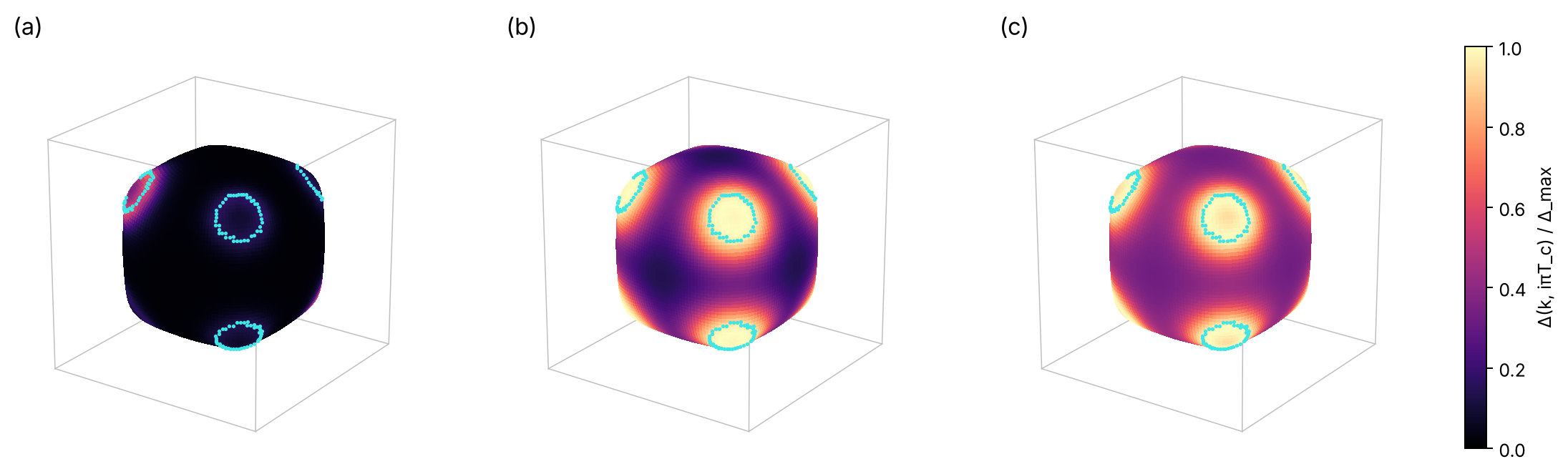}
\caption{Gap $\Delta(\kk,i\pi T_c)$, normalized to its maximum, on the $n=0.6$ Fermi surface from Eq.~\eqref{eq:eliash} with $\langle\lambda_R\rangle=0.70$ and $\omega_m=25$~meV. (a) $R$ boson alone, $\Delta q=0.03$ ($2\pi/a$), $\mu^*=0$; $T_c=45.7$~K, $\Delta_{\min}/\Delta_{\max}=0.008$. (b) $R$ boson alone, $\Delta q=0.1$ ($2\pi/a$), $\mu^*=0.1$; $T_c=11.7$~K, $\Delta_{\min}/\Delta_{\max}=0.14$. (c) $R$ boson plus the 50~meV phonon, $\Delta q=0.07$ ($2\pi/a$), $\mu^*=0.1$; $T_c=26.6$~K, $\Delta_{\min}/\Delta_{\max}=0.33$. Cyan: hot lines on the visible side.}
\label{fig:gap}
\end{figure*}

The gap follows the coupling [Fig.~\ref{fig:gap}(a,b)]. For $\Delta q=0.03$ and $\mu^*=0$ it is concentrated on the hot loops, where it is more than 100 times larger than on the cold faces of the Fermi surface. For $\Delta q=0.1$ it is still maximal on the loops, with $\Delta_{\min}/\Delta_{\max}=0.14$ along $\langle 100\rangle$. For narrow peaks the four pairs of hot loops, each loop paired by $\QQ$ with the loop around the opposite $\langle 111\rangle$ direction, are only weakly coupled to each other. The Coulomb pseudopotential then favors solutions in which the gap changes sign between loop pairs, and for $\Delta q\le 0.05$ with $\mu^*=0.1$ such a solution is nearly degenerate with the uniform-sign one, at $T_c$ reduced by only 1~K relative to $\mu^*=0$. We analyze this competition by symmetry in Sec.~\ref{sec:sym}.

\subsection{$R$ boson plus breathing-mode phonon}

Adding the isotropic phonon channel ($\lambda_b=0.3$ in the 50~meV Gaussian of Fig.~\ref{fig:a2F}) raises the uniform-coupling $T_c$ from 6 to 20~K and makes the gap more uniform, because the phonon couples all parts of the Fermi surface equally. Table~\ref{tab:aniso} compares the two cases. Replacing the Gaussian by an Einstein mode at 50~meV changes every $T_c$ in the table by less than 0.2~K, so the results do not depend on the width of the phonon peak.

\begin{table*}[t]
\caption{Anisotropic Eliashberg solution of Eq.~\eqref{eq:eliash} with $\langle\lambda_R\rangle=0.70$, $\omega_m=25$~meV and $\mu^*=0.1$. $\Delta_{\min}/\Delta_{\max}$ is the gap ratio on the Fermi surface at $T_c$; a negative entry indicates a sign change between hot-loop pairs.}
\label{tab:aniso}
\begin{ruledtabular}
\begin{tabular}{ccccccc}
\shortstack{Peak width\\$\Delta q$ ($2\pi/a$)} & \shortstack{Correlation\\length $\xi_c/a$} & \shortstack{Range of\\$\lambda_R(\kk)$ on FS} & \shortstack{$T_c$ (K),\\boson only} & \shortstack{$\Delta_{\min}/\Delta_{\max}$,\\boson only} & \shortstack{$T_c$ (K),\\boson + phonon} & \shortstack{$\Delta_{\min}/\Delta_{\max}$,\\boson + phonon}\\
\hline
uniform & -- & 0.70 & 6.1 & 1 & 20.3 & 1\\
0.2 & 0.8 & 0.45--0.96 & 6.8 & 0.52 & 20.7 & 0.75\\
0.1 & 1.6 & 0.24--1.42 & 11.7 & 0.14 & 23.1 & 0.46\\
0.07 & 2.3 & 0.17--1.84 & -- & -- & 26.6 & 0.33\\
0.05 & 3.2 & 0.12--2.41 & 28.2 & $<0$ & 31.5 & 0.23\\
0.03 & 5.3 & 0.07--4.42 & 44.9 & $<0$ & 42.0 & 0.08\\
\end{tabular}
\end{ruledtabular}
\end{table*}

Two conclusions follow. First, with the total coupling fixed at $\lambda\approx 1$, $T_c\approx 25$--30~K requires $\Delta q\approx 0.05$--0.07 ($2\pi/a$), corresponding to $\xi_c\approx 2$--3 lattice spacings. A sharper peak concentrates too much coupling on the hot lines and gives $T_c\approx 40$~K, while a broader one gives at most 23~K. Second, in this regime the gap is strongly anisotropic [Fig.~\ref{fig:gap}(c)]: it is largest on and around the hot loops near $\langle 111\rangle$ and smallest along $\langle 100\rangle$, with $\Delta_{\langle 111\rangle}/\Delta_{\langle 100\rangle}\approx 3$--4. The quasiparticle mass follows the same pattern, $Z^{-1}\approx 3.1$ near the loops against 1.5 along $\langle 100\rangle$ for $\Delta q=0.07$. This anisotropy is large enough to detect by angle-resolved photoemission, but only if the hot regions near $\langle 111\rangle$ are sampled; cuts along $\langle 100\rangle$ would see a nearly isotropic, smaller gap. The ARPES study of Ref.~\cite{Wen2018}, which found an isotropic gap, was performed on Ba$_{0.51}$K$_{0.49}$BiO$_3$. Besides possibly not sampling the hot regions at the appropriate $k_z$, this composition lies further from the charge-ordered phase than $x=0.4$, which should weaken the $R$-point fluctuations and reduce the anisotropy.

The uniform two-channel $T_c$ of 20~K lies below the observed 30~K; the concentration of the coupling on the hot loops supplies the remaining enhancement. Since $\lambda_b=0.3$ at 50~meV places more weight at high frequency than the tunneling spectrum shows, reducing the phonon weight to the measured value would shift the consistent range of $\Delta q$ slightly toward narrower peaks. The consistent regime is also where the pseudogap criterion~\eqref{eq:pgcrit} is far from satisfied, so the normal state remains a Fermi liquid with moderately heavy quasiparticles on the hot bands, as assumed.

\subsection{Competing gap symmetries on the hot loops}
\label{sec:sym}

The eight hot loops lie around the four body diagonals, and $\QQ$ connects each loop to the one around the opposite direction. For a narrow peak the $R$ boson pairs electrons within each such loop pair but couples different pairs only weakly. A singlet gap concentrated on the loops is even under inversion and is therefore specified by four amplitudes, one per body diagonal. Under the cubic group these span two representations: $A_{1g}$, with the same sign on all four pairs, and $T_{2g}$, whose three partners have the sign structure of $k_xk_y$, $k_yk_z$ and $k_zk_x$ evaluated on the diagonals. In the $T_{2g}$ state the gap changes sign between loop pairs, and the sign changes lie on the cold parts of the Fermi surface, where the coupling is weak.

The two states respond oppositely to the isotropic parts of the interaction. Because the $T_{2g}$ gap averages to zero over the Fermi surface, it is untouched by the Coulomb pseudopotential and receives no pairing from the isotropic phonon channel, only its mass renormalization. The $A_{1g}$ gap gains from the phonon and loses to $\mu^*$. To compare them we solved Eq.~\eqref{eq:eliash} separately in each symmetry sector, on a Fermi-surface mesh of 1056 points generated from the irreducible wedge so that the cubic group acts exactly, projecting the kernel onto each representation.

\begin{table}[t]
\caption{$T_c$ (K) of the leading $A_{1g}$ and $T_{2g}$ solutions, $\langle\lambda_R\rangle=0.70$, $\omega_m=25$~meV. Bold: the state with higher $T_c$.}
\label{tab:sym}
\begin{ruledtabular}
\begin{tabular}{ccccc}
\shortstack{$\Delta q$\\($2\pi/a$)} & \shortstack{Phonon\\$\lambda_b$} & \shortstack{Coulomb\\$\mu^*$} & \shortstack{$T_c$ (K)\\$A_{1g}$ state} & \shortstack{$T_c$ (K)\\$T_{2g}$ state}\\
\hline
0.03 & 0 & 0 & \textbf{47.6} & 46.9\\
0.03 & 0 & 0.1 & 43.9 & \textbf{46.9}\\
0.03 & 0 & 0.2 & 41.6 & \textbf{46.9}\\
0.05 & 0 & 0.1 & 27.6 & \textbf{29.5}\\
0.1 & 0 & 0.1 & \textbf{12.1} & 10.3\\
0.03 & 0.3 & 0.1 & \textbf{43.9} & 37.9\\
0.05 & 0.3 & 0.1 & \textbf{32.4} & 22.6\\
0.1 & 0.3 & 0.1 & \textbf{23.5} & 6.8\\
\end{tabular}
\end{ruledtabular}
\end{table}

\begin{figure*}[t]
\includegraphics[width=\textwidth]{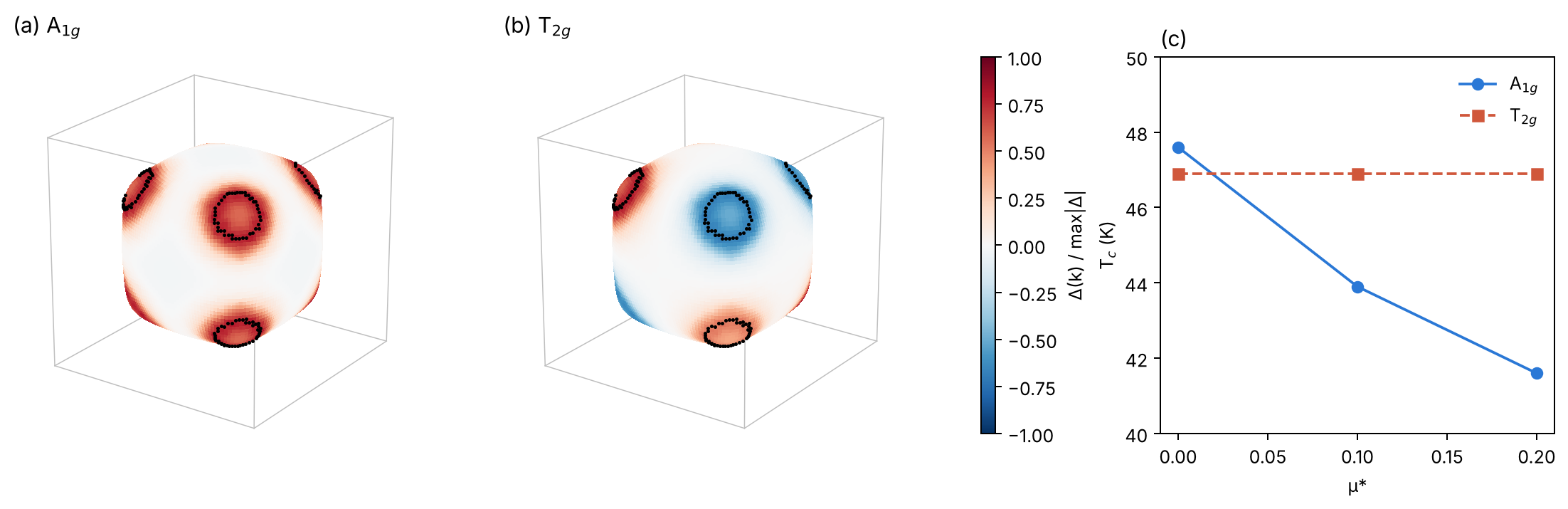}
\caption{Leading gap functions in the two symmetry sectors for the $R$ boson alone, $\Delta q=0.03$ ($2\pi/a$), $\mu^*=0.1$, normalized to $\max|\Delta|$. (a) $A_{1g}$: the same sign on all hot loops, $T_c=43.9$~K. (b) $T_{2g}$: one partner of the triplet, with the gap changing sign between loop pairs around different body diagonals, $T_c=46.9$~K. Black: hot lines. (c) $T_c$ of both states versus $\mu^*$ for $\Delta q=0.03$ ($2\pi/a$); the $T_{2g}$ $T_c$ is independent of $\mu^*$.}
\label{fig:sym}
\end{figure*}

With the $R$ boson alone and a narrow peak, the two states are within a few kelvin of each other, and the sign-changing $T_{2g}$ state wins once $\mu^*$ exceeds about 0.02 [Table~\ref{tab:sym} and Fig.~\ref{fig:sym}(c)]. Its $T_c$ is independent of $\mu^*$, as expected. Broadening the peak to $\Delta q=0.1$ couples the loop pairs through the tails of $V(\qq)$ and favors $A_{1g}$ by 2~K. Adding the isotropic phonon channel favors $A_{1g}$ decisively, by 6~K for $\Delta q=0.03$, 10~K for $\Delta q=0.05$ and 17~K for $\Delta q=0.1$; in the regime consistent with $T_c$ and tunneling the margin is about 10~K. The $A_{1g}$ gap itself acquires a weak opposite sign on the coldest parts of the Fermi surface for narrow peaks without the phonon ($\Delta_{\min}/\Delta_{\max}\approx-0.01$ to $-0.06$), a mild $s_\pm$ structure between hot loops and cold faces driven by $\mu^*$.

Which state is realized therefore depends on three quantities that are not fixed by the tunneling spectrum alone: the correlation length $\xi_c$, the strength of the isotropic phonon coupling, which the tunneling data suggest may be smaller than the LDA value, and $\mu^*$ on the hot loops. Because both states place essentially all of their gap on the hot loops, their gap magnitudes are nearly identical there, and photoemission measuring $|\Delta(\kk)|$ near $\langle 111\rangle$ cannot distinguish them. Several experiments can. (i)~\textit{Phase-sensitive tests.} The $T_{2g}$ gap changes sign under a $90^\circ$ rotation about a cubic axis, which can be probed by corner-junction or tricrystal Josephson experiments on cubic BKBO, by analogy with the tests of $d$-wave pairing in the cuprates. (ii)~\textit{Disorder.} BKBO is an intrinsically disordered alloy. For a gap that averages to zero over the Fermi surface, such as $T_{2g}$, scattering by nonmagnetic impurities is pair breaking, and $T_c$ is suppressed according to the Abrikosov--Gor'kov form $\ln(T_{c0}/T_c)=\psi\!\left(\tfrac12+\hbar/4\pi k_BT_c\tau\right)-\psi\!\left(\tfrac12\right)$, where $\tau$ is the elastic scattering time and $\psi$ the digamma function~\cite{Abrikosov1960,Balatsky2006}; superconductivity disappears once $\hbar/\tau$ exceeds about $1.8\,k_BT_{c0}$. A same-sign $A_{1g}$ gap is instead protected by Anderson's theorem~\cite{Anderson1959}. A systematic study of $T_c$ against residual resistivity, or against controlled irradiation, discriminates the two. (iii)~\textit{Low-energy excitations.} The $T_{2g}$ state has lines of gap nodes on the cold parts of the Fermi surface, giving power-law contributions to the specific heat, penetration depth and thermal conductivity. Because the gap on the cold regions is small in both states, these signatures would appear at temperatures well below $T_c$. Existing tunneling spectra show a fully developed gap with low zero-bias conductance~\cite{Huang1990,Samuely1993}, which favors $A_{1g}$, but the small cold-region weight makes this test less sharp than in the cuprates. (iv)~\textit{Momentum-resolved phase.} Phase-referenced quasiparticle interference in scanning tunneling spectroscopy, or the presence or absence of a resonance-like feature in the charge response at $\QQ$ below $T_c$, would also distinguish the sign structure.

On present evidence the $A_{1g}$ state is favored: it wins in the parameter range consistent with $T_c$ and the tunneling spectrum, and it is compatible with the full gap and with superconductivity surviving strong alloy disorder. The $T_{2g}$ state remains a close competitor if the isotropic phonon coupling is weaker or the charge correlations longer-ranged than we estimate, and only measurements sensitive to the sign of the gap on the hot loops can settle the question.

\section{Toward a microscopic picture: random-phase approximation}
\label{sec:rpa}

The phenomenological results of Secs.~\ref{sec:normal} and \ref{sec:sc} do not depend on the microscopic origin of the boson. They rest on the overdamped boson model of Sec.~\ref{sec:normal}, a charge mode at $\QQ=R$ with three physical parameters: (1)~its strength $\Ub$, fixed through the sum rule by the coupling $\lambda_R\approx 0.7$ missing from the LDA, which corresponds to $\Ub\approx 3$~eV; (2)~its frequency $\omega_m\approx 25$~meV, which places the coupling in the measured 10--50~meV range; and (3)~its momentum width $\Delta q=1/\xi_c$, the one parameter adjusted to $T_c$, which gives $\xi_c\approx 2$--3$a$. The ordering vector $\QQ$ is not free, since it is set by the charge order of the parent compound, and the functional form of $V(\qq)$ and the high-frequency cutoff are modeling choices to which the results are only weakly sensitive. A microscopic theory should account for these three numbers. In this section we ask how far a minimal model, the random-phase approximation for an extended Hubbard model, goes toward that goal.

\subsection{Extended Hubbard model and the static instability}

For a single band, the charge response can be computed directly in the random-phase approximation. We use the tight-binding model described in Appendix~\ref{app:tb}, supplemented by an on-site repulsion $U_0$ and a nearest-neighbor repulsion $V_0$,
\begin{equation}
\begin{split}
U(\qq)&=U_0+V(\qq),\\ V(\qq)&=2V_0\left(\cos q_x+\cos q_y+\cos q_z\right),
\end{split}
\label{eq:Uq}
\end{equation}
with $U_0=0.6$~eV. Since $U_0$ acts only between opposite spins while $V(\qq)$ acts between all pairs, the charge susceptibility in terms of the per-spin Lindhard function $\chi_0^\sigma$ is
\begin{equation}
\chi_c(\qq,\omega)=\frac{2\chi_0^\sigma(\qq,\omega)}{1+K(\qq)\,\chi_0^\sigma(\qq,\omega)},\qquad K(\qq)=U_0+2V(\qq),
\label{eq:chic}
\end{equation}
with
\begin{equation}
\chi_0^\sigma(\qq,\omega)=\frac{1}{N}\sum_\kk\frac{f(\xi_\kk)-f(\xi_{\kk+\qq})}{\xi_{\kk+\qq}-\xi_\kk-\omega-i0^+}.
\label{eq:chi0}
\end{equation}
The static $\chi_0^\sigma(\qq)$ at $n=0.6$ [Fig.~\ref{fig:rpa}(a)] equals $N_\sigma(0)=0.22$~eV$^{-1}$ at $\Gamma$, rises to 0.26~eV$^{-1}$ at $X$ and reaches its maximum, 0.34~eV$^{-1}$, at $M$; at $R$ it is 0.32~eV$^{-1}$. Although $\chi_0^\sigma$ is largest at $M$, the kernel $K(\qq)$ is most negative at $R$, where $V(R)=-6V_0$. The instability, $1+K(\qq)\chi_0^\sigma(\qq)=0$, therefore occurs first at $\QQ=R$, at
\begin{equation}
V_0^c=\frac{U_0+1/\chi_0^\sigma(\RR)}{12}\approx 0.31~\text{eV}.
\label{eq:V0c}
\end{equation}
Neighboring wave vectors become unstable only at higher $V_0$, by 3--7\% for $\qq$ displaced by $\pi/8a$ from $R$ along one or two axes. The threshold depends little on temperature (0.311--0.313~eV for $k_BT=2.5$--20~meV) and on the integration method; a tetrahedron calculation gives 0.295~eV. Near $R$, $\chi_0^\sigma$ is flat to better than 1\% over $0.3/a$, so the instability is driven by the attractive kernel at $R$ rather than by nesting. This is the commensurate breathing-type charge order of the parent compound, here reached from the metallic side.

\subsection{Dynamics: RPA produces the overdamped boson}

At the ordering vector the bare susceptibility has a particularly simple frequency dependence. Over 0--150~meV,
\begin{equation}
\chi_0^\sigma(\RR,\omega)\simeq\chi_0'+i\gamma\omega,
\label{eq:chiR}
\end{equation}
with $\chi_0'=0.316$~eV$^{-1}$ and $\gamma=\pi J_\QQ(0,0)=0.141$~eV$^{-2}$, the real part constant to 0.3\% and $\chi_0''/\omega$ constant to a few percent. Here $J_\QQ(0,0)=N^{-1}\sum_\kk\delta(\xi_\kk)\delta(\xi_{\kk+\QQ})$ is the joint density of states on the hot lines: the Landau damping of the $R$ mode comes entirely from particle--hole pairs on the same lines that carry the mass enhancement of Sec.~\ref{sec:normal}. Its evaluation requires care, since a $\kk$-mesh commensurate with $\QQ$ systematically misses the hot lines; we used Monte Carlo sampling with $8\times 10^6$ random points and obtained the real part by Kramers--Kronig transformation.

\begin{figure*}[t]
\includegraphics[width=\textwidth]{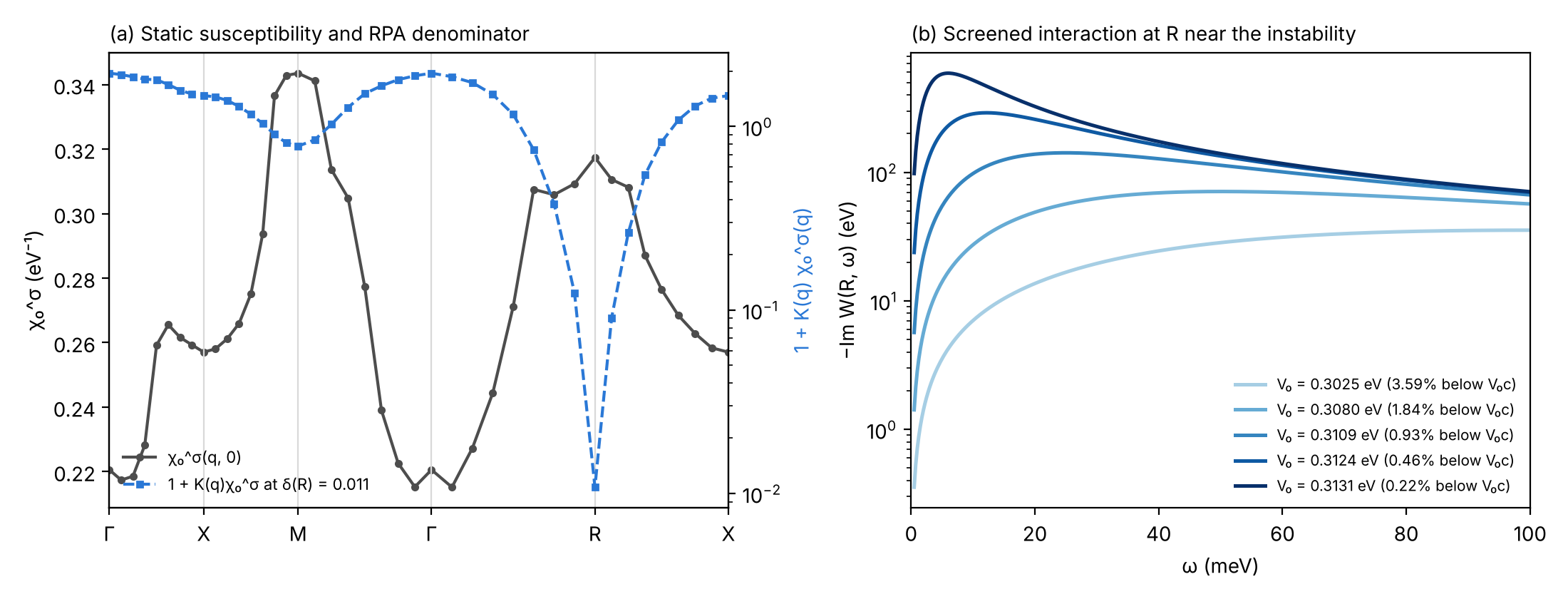}
\caption{(a) Static per-spin Lindhard function $\chi_0^\sigma(\qq)$ at $n=0.6$ along high-symmetry lines (gray, left axis) and the RPA denominator $1+K(\qq)\chi_0^\sigma(\qq)$ for $V_0$ tuned to $\delta(R)=0.011$ (blue, right axis, logarithmic). The susceptibility peaks at $M$, but the instability occurs at $R$. (b) Spectral function $-\mathrm{Im}[K/(1+K\chi_0^\sigma)]=-2\,\mathrm{Im}\,W_c$ at $R$ for $V_0$ approaching $V_0^c=0.314$~eV, using Eq.~\eqref{eq:chiR}; the peak sits at $\omega_m=\delta/(|K|\gamma)$.}
\label{fig:rpa}
\end{figure*}

Inserting Eq.~\eqref{eq:chiR} into Eq.~\eqref{eq:chic}, the charge-channel interaction acting on an electron, $W_c=(K/2)/(1+K\chi_0^\sigma)$, becomes
\begin{equation}
W_c(\RR,\omega)=-\frac{1/2\gamma}{\omega_m-i\omega},\qquad\omega_m=\frac{\delta}{|K(\RR)|\,\gamma},
\label{eq:WcR}
\end{equation}
where $\delta=1+K(\RR)\,\chi_0'$. This is exactly the relaxational form~\eqref{eq:W}, with a prefactor independent of $V_0$ and a characteristic frequency proportional to the distance $\delta$ from the instability. The phenomenological boson of Secs.~\ref{sec:normal} and \ref{sec:sc} is therefore the RPA charge fluctuation at $R$, overdamped by the hot lines. As $V_0$ approaches $V_0^c$ the peak of $-\mathrm{Im}\,W$ moves down linearly [Fig.~\ref{fig:rpa}(b)]: $\omega_m\approx 100$, 48, 26, 12 and 6~meV for $V_0$ 3.6, 1.8, 0.93, 0.46 and 0.22\% below $V_0^c$. The frequency $\omega_m\approx 25$~meV adopted in Sec.~\ref{sec:sc} requires $\delta\approx 0.011$, that is $V_0$ within about 1\% of the threshold and a static enhancement of the charge response by a factor of about 90. Because $\omega_m$ depends only on $\delta$, this conclusion does not depend on the method used to compute $V_0^c$.

\subsection{Can electronic RPA account for the phenomenology?}
\label{sec:rpawidth}

RPA reproduces the form of the boson, but not its momentum width or its coupling strength.

\textit{Width.} Near the instability the static interaction has the Ornstein--Zernike form~\cite{Ornstein1914}, $W_c(\RR+\mathbf{p},0)\propto 1/(\delta+cp^2)$, with a half-width that scales as $\sqrt\delta$. At $\delta=0.011$, where $\omega_m\approx 25$~meV, the half-width is about $0.20/a$ [$0.032$ ($2\pi/a$)] along $[001]$ and $0.12/a$ along $[111]$, so $\xi_c\approx 5$--9 lattice spacings. In RPA the frequency and width are tied by the dynamical exponent $z=2$, $\omega_m\propto\Delta q^2$. The broader peak, $\Delta q\approx 0.05$--0.07 ($2\pi/a$), required in Sec.~\ref{sec:sc} to reproduce $T_c$ would correspond to $\delta\approx 0.03$--0.05 and $\omega_m\approx 60$--120~meV, several times the boson scale. Electronic RPA cannot produce a boson that is both soft and short-ranged.

\textit{Strength.} The Fermi-surface-averaged coupling follows from the sum rule~\eqref{eq:sumrule}, now applied to the dynamically screened part of the RPA interaction, $\Delta W_c=W_c(\qq,0)-K(\qq)/2$:
\begin{equation}
\lambda_R^{\rm RPA}\simeq\frac{J_\QQ}{N_\sigma}\int\frac{d^3q}{(2\pi)^3}\,|\Delta W_c(\qq,0)|,\qquad\frac{J_\QQ}{N_\sigma}\approx 0.22~\text{eV}^{-1}.
\label{eq:lamRPA}
\end{equation}
At $\delta=0.011$ the peak around $R$ contributes $\lambda_R^{\rm RPA}\approx 0.04$, and the rest of the zone at most about 0.08. The effective $\Ub$ is therefore about 0.2~eV, compared with the $\approx 3$~eV implied by $\lambda_R=0.7$. Approaching the instability does not help: in three dimensions the Ornstein--Zernike integral $\int d^3p/(\delta+cp^2)$ remains finite as $\delta\to 0$, so the peak grows taller and narrower while its weight in Eq.~\eqref{eq:lamRPA} saturates. Closer to threshold only $\omega_m$ decreases.

The results of this subsection can be summarized against the three parameters of the overdamped boson model listed at the beginning of this section. Electronic RPA accounts for the structure of the boson but fixes only one of its three parameters:
\begin{itemize}
\item \textit{Ordering vector and form.} RPA reproduces both: the instability occurs first at $\QQ=R$, and near it the screened interaction has exactly the overdamped form of Eq.~\eqref{eq:W}, with the Landau damping $\gamma$ coming from the hot lines.
\item \textit{(2) Frequency $\omega_m$.} RPA can produce $\omega_m\approx 25$~meV, but only by tuning $V_0$ to within about 1\% of the instability, since $\omega_m$ is proportional to the distance $\delta$ from it.
\item \textit{(3) Correlation length $\xi_c$.} RPA cannot produce it independently. The dynamical exponent $z=2$ ties it to the frequency, $\omega_m\propto\Delta q^2$: at $\omega_m\approx 25$~meV it gives $\xi_c\approx 5$--9$a$, whereas the $\xi_c\approx 2$--3$a$ required by $T_c$ would correspond to $\omega_m\approx 60$--120~meV.
\item \textit{(1) Strength $\Ub$.} RPA falls short by roughly an order of magnitude: $\Ub\approx 0.2$~eV and $\lambda_R\approx 0.04$--0.1, against $\Ub\approx 3$~eV and $\lambda_R\approx 0.7$, and in three dimensions moving closer to the instability does not help.
\end{itemize}
An additional ingredient is therefore needed to enhance the vertex between electrons and the $R$-point charge fluctuation, addressing (1), and to decouple its frequency from its correlation length, addressing (3).

\subsection{Including the breathing-mode electron--phonon coupling}

One natural candidate is the lattice. In BaBiO$_3$ the $R$-point charge order is a breathing distortion of the oxygen octahedra, so the charge fluctuation and the bond-stretching modes are strongly mixed. We extract the electron--phonon vertex from the mode-resolved LDA couplings of Meregalli and Savrasov~\cite{Meregalli1998}. For each wave vector on their $6\times6\times6$ mesh, the Fermi-surface-averaged vertex of the breathing branch is
\begin{equation}
g^2(\qq)=\frac{\lambda_{\qq\nu}\,\omega_{\qq\nu}\,N_\sigma}{2J(\qq)},\qquad W_{\rm ph}(\qq,0)=-\frac{2g^2(\qq)}{\omega_{\qq\nu}},
\label{eq:g2}
\end{equation}
with $J(\qq)$ the nesting function of our band. The result follows the bond-stretching form $g^2(\qq)\approx g_0^2\sum_i\sin^2(\pi q_i)$, with $g_0^2\approx 0.02$~eV$^2$, to within about 30\% except along $R$--$X$, where it is about half as large. At $R$, $g^2=0.054$~eV$^2$ and the static phonon-mediated attraction is $W_{\rm ph}\approx-1.7$~eV, comparable to the electronic charge-channel kernel $K(\RR)/2\approx-1.6$~eV near threshold. The lattice is therefore not a small correction at $R$.

Formally, the phonon adds a retarded, phonon-mediated interaction $V_{\rm ph}(\qq,\omega)=g^2(\qq)D(\qq,\omega)$, with phonon propagator $D(\qq,\omega)=2\Omega_\qq/(\omega^2-\Omega_\qq^2)$, which acts between all electrons like $V(\qq)$. The RPA charge kernel and the screened interaction of Eq.~\eqref{eq:chic} then become
\begin{equation}
\begin{split}
K(\qq,\omega)&=U_0+2V(\qq)+2g^2(\qq)\,\frac{2\Omega_\qq}{\omega^2-\Omega_\qq^2},\\
W_c(\qq,\omega)&=\frac{K(\qq,\omega)/2}{1+K(\qq,\omega)\,\chi_0^\sigma(\qq,\omega)}.
\end{split}
\label{eq:Kph}
\end{equation}
The poles of $W_c$ describe the coupled charge--lattice modes; at $\omega=0$ the phonon term reduces to $-2g^2/\Omega_\qq$, that is $2W_{\rm ph}$. Equation~\eqref{eq:Kph} holds for bare phonons and a bare vertex, which are not what the LDA provides.

The LDA frequencies and couplings are already screened by the LDA kernel $K_{\rm LDA}(\qq)$, so they cannot be added to the bare RPA interaction without double counting. Using the LDA phonons as the reference, only the interaction beyond LDA, $\Delta K(\qq)=K(\qq)-K_{\rm LDA}(\qq)$, renormalizes them:
\begin{equation}
\begin{split}
\Omega_{\rm eff}^2(\qq)&=\Omega_{\rm LDA}^2+\frac{4\,\Omega_{\rm LDA}\,g^2\,\Delta K\,(\chi_0^\sigma)^2}{1+\Delta K\,\chi_0^\sigma},\\ \frac{\lambda_{\rm eff}}{\lambda_{\rm LDA}}&=\frac{\Omega_{\rm LDA}}{\Omega_{\rm eff}}\,\frac{1}{(1+\Delta K\,\chi_0^\sigma)^2}.
\end{split}
\label{eq:soft}
\end{equation}
At $R$, with $\Omega_{\rm LDA}=65$~meV, an attractive $\Delta K(R)=-0.5$~eV softens the breathing mode by 10\% to 58~meV and raises its coupling from 0.32 to 0.51; $\Delta K=-1$~eV gives 47~meV and 0.96. The coupled lattice--charge instability occurs at $\Delta K\approx-1.55$~eV, only half the purely electronic threshold $|K|=1/\chi_0^\sigma\approx 3.2$~eV. If $K_{\rm LDA}(R)\approx 0$, this corresponds to $V_0\approx 0.18$~eV instead of 0.31~eV. Moderate correlations beyond LDA can therefore correct the overestimated LDA breathing frequency and enhance the coupling at the same time, the RPA counterpart of the correlation-enhanced electron--phonon coupling of Refs.~\cite{Yin2013,Li2019}.

Near the coupled instability the soft mode inherits a strong vertex from the phonon and a frequency set partly by the lattice rather than by $\delta$ alone. This could supply the ingredient missing from purely electronic RPA: it may raise the coupling to the level required by tunneling, spread it over the 10--30~meV range where the oxygen and Bi modes lie, which would account for the peaks in the measured $\alpha^2F(\omega)$, and relax the $z=2$ relation between $\omega_m$ and $\xi_c$. A quantitative treatment requires the momentum dependence of $K_{\rm LDA}(\qq)$, which we do not have for this band, and a full coupled-mode calculation over the Brillouin zone. We leave this to future work, and note that it is the most direct route to fixing the two phenomenological parameters of Secs.~\ref{sec:normal} and \ref{sec:sc}, $\Ub$ and $\xi_c$, from first principles.

\section{Conclusion}
\label{sec:concl}

We have proposed that the electron--boson coupling missing from density-functional calculations of Ba$_{1-x}$K$_x$BiO$_3$ comes from dynamical charge fluctuations at $\QQ=(\pi,\pi,\pi)/a$, the remnant in the metal of the parent compound's breathing-type charge order. We describe them by an overdamped boson model with three physical parameters: the strength of the interaction, its characteristic frequency $\omega_m$, and its correlation length $\xi_c$. Such a boson acts almost exclusively on hot lines of the Fermi surface, small loops around $\langle 111\rangle$ where $\kk$ and $\kk+\QQ$ both lie at the Fermi level. There it produces heavy quasiparticles and, for long correlation lengths, shadow bands and a pseudogap, while a sum rule fixes the Fermi-surface-averaged coupling at $\Ub$ times the nesting density, independently of the boson frequency and correlation length.

Taking the total coupling from tunneling as $\lambda\approx 1$, the LDA breathing mode supplies about 0.3 and the overdamped boson the missing 0.7, with a frequency $\omega_m\approx 25$~meV compatible with $\omega_{\log}\approx 30$--40~meV and with a linear low-frequency onset that the tunneling inversion cannot exclude. Anisotropic Eliashberg calculations then reproduce $T_c\approx 25$--30~K for correlation lengths of two to three lattice spacings. In this regime the gap is $s$-wave but enhanced by a factor of 3--4 near $\langle 111\rangle$, and the quasiparticles there are about twice as heavy as along $\langle 100\rangle$. For a sharp peak the problem reduces to independent pairs on the hot loops with $T_c\propto\sqrt{\Ub\omega_m}$, and a sign-changing $T_{2g}$ state competes closely with $s$-wave pairing.

In the random-phase approximation for an extended Hubbard model the overdamped boson model emerges naturally near a charge instability at $R$, with exactly the assumed form and with damping by the hot lines. Its frequency can be matched only by tuning close to the instability, however, and purely electronic fluctuations are an order of magnitude too weak and too long-ranged to account for the other two parameters. Their coupling to the breathing-mode lattice distortion, which softens the mode and enhances the vertex, is one plausible source of the missing strength, which a full coupled-mode calculation would need to confirm. The theory makes direct experimental predictions: a gap and mass enhancement concentrated near the $\langle 111\rangle$ directions, observable by angle-resolved photoemission at suitable $k_z$; a soft, broad charge response at the $R$ point in inelastic x-ray or neutron scattering; and, if the $T_{2g}$ state is realized, signatures of sign changes in phase-sensitive and disorder experiments.

\appendix
\setcounter{figure}{0}
\renewcommand{\thefigure}{A\arabic{figure}}
\section{Tight-binding fit of the conduction band}
\label{app:tb}

The conduction band is represented by a Fourier series on the simple-cubic lattice,
\begin{equation}
\varepsilon(\kk)=\sum_\RR t(\RR)\,e^{i\kk\cdot\RR}=\sum_s t_s\sum_{\RR\in s}\cos(\kk\cdot\RR),
\label{eq:A1}
\end{equation}
where $s$ runs over the 14 shells of symmetry-equivalent lattice vectors $\RR=(l,m,n)a$ listed in Table~\ref{tab:tb}, including the on-site term. The parameters were fitted to the conduction band of cubic BaBiO$_3$ found from density-functional calculations in the LDA, shown in Fig.~\ref{fig:lda}; doping is treated in the rigid-band approximation by fixing the electron count. Energies are measured from the half-filling energy.

\begin{table}[t]
\caption{Tight-binding parameters of Eq.~\eqref{eq:A1}.}
\label{tab:tb}
\begin{ruledtabular}
\begin{tabular}{ccr}
Shell $(l\,m\,n)$ & Vectors in shell & $t$ (Ry)\\
\hline
0 0 0 & 1 & $-0.004276$\\
1 0 0 & 6 & $-0.033732$\\
1 1 0 & 12 & $-0.004074$\\
1 1 1 & 8 & $0.001826$\\
2 0 0 & 6 & $0.010065$\\
2 1 0 & 24 & $0.001172$\\
2 1 1 & 24 & $-0.000648$\\
2 2 0 & 12 & $-0.000801$\\
2 2 1 & 24 & $0.000808$\\
3 0 0 & 6 & $-0.002495$\\
3 1 0 & 24 & $0.001248$\\
3 1 1 & 24 & $-0.000405$\\
2 2 2 & 8 & $-0.000448$\\
3 2 0 & 24 & $-0.000366$\\
\end{tabular}
\end{ruledtabular}
\end{table}

The band velocity is $\mathbf{v}(\kk)=-\sum_\RR t(\RR)\,\RR\sin(\kk\cdot\RR)$. The band spans $-0.169$ to $+0.190$~Ry (bandwidth 4.89~eV), with $\varepsilon=-0.165$~Ry at $\Gamma$, $-0.010$~Ry at $X$, $+0.119$~Ry at $M$ and $+0.182$~Ry at $R$. The minimum lies slightly off $\Gamma$ along $\Gamma$--$R$ and the maximum partway along $M$--$R$, because the $(2,0,0)$ hopping is positive. The saddle point at $X$ lies 0.14~eV below the half-filling energy, giving a density of states at half filling of 5.5 states/(Ry$\cdot$spin$\cdot$cell).

\begin{figure}[t]
\includegraphics[width=\columnwidth]{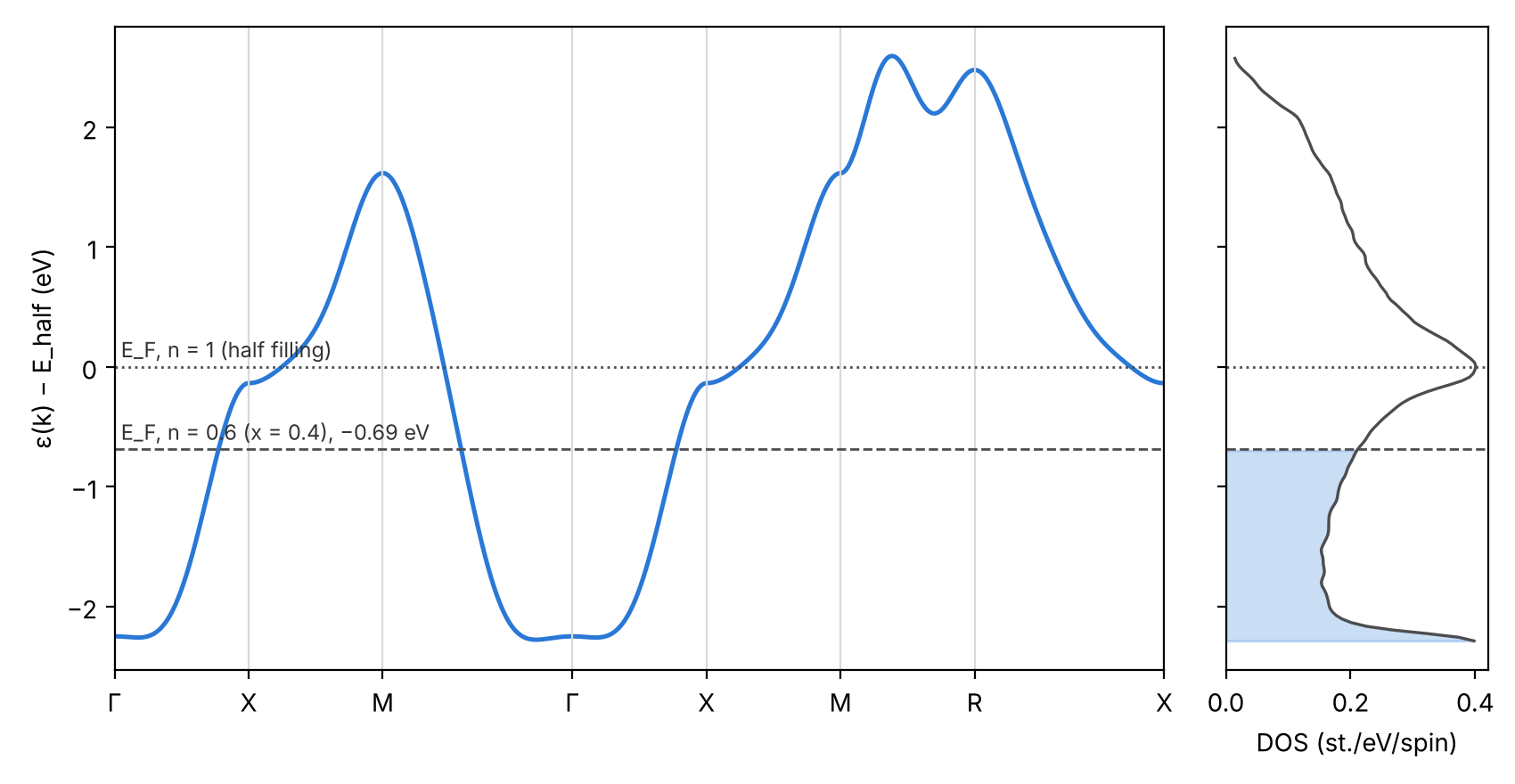}
\caption{Tight-binding conduction band of Eq.~\eqref{eq:A1} along the same path as the LDA bands of Fig.~\ref{fig:lda}, $\Gamma$--$X$--$M$--$\Gamma$--$X$--$M$--$R$--$X$ (left) and its density of states per spin (right), with energies measured from the half-filling energy. Dotted: Fermi level for $n=1$ (half filling, parent BaBiO$_3$), lying 0.14~eV above the van Hove saddle point at $X$. Dashed: Fermi level for $n=0.6$ ($x=0.4$), $\mu=-0.69$~eV, used throughout this paper; the shaded area is the occupied part of the band.}
\label{fig:tb}
\end{figure}

All calculations in this paper use $n=0.6$ electrons per site ($x=0.4$), for which $\mu=-0.0506$~Ry ($-0.69$~eV) relative to half filling and $N_\sigma(0)=2.9$ states/(Ry$\cdot$spin$\cdot$cell) $=0.21$~eV$^{-1}$. The band and its density of states are shown in Fig.~\ref{fig:tb}.

\end{document}